\documentclass{article}

\usepackage{arxiv}

\usepackage[utf8]{inputenc} 
\usepackage[T1]{fontenc}    
\usepackage{hyperref}       
\usepackage{url}            
\usepackage{booktabs}       
\usepackage{amsfonts}       
\usepackage{nicefrac}       
\usepackage{microtype}      
\usepackage{lipsum}		    
\usepackage{graphicx}
\usepackage{natbib}
\usepackage{doi}
\usepackage[svgnames]{xcolor}
\usepackage{listings}       

\title{\emph{FMM}: An R Package for Modeling Rhythmic Patterns in Oscillatory Systems}

\author{ \href{https://orcid.org/0000-0002-5077-4448}{\includegraphics[scale=0.06]{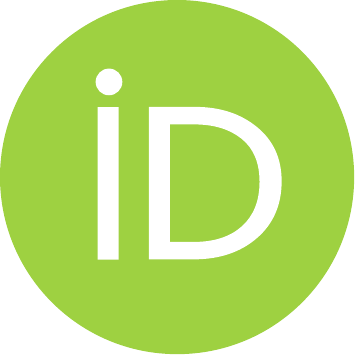}\hspace{1mm}Itziar Fern\'andez}\\
	Department of Statistics and Operations Research\\
 	Universidad of Valladolid, Valladolid, Spain \\
	\texttt{itziar.fernandez@uva.es} \\
	\And
	\href{https://orcid.org/0000-0001-5450-9580}{\includegraphics[scale=0.06]{orcid.pdf}\hspace{1mm}Alejandro Rodr\'iguez-Collado} \\
	Department of Statistics and Operations Research\\
 	Universidad of Valladolid, Valladolid, Spain \\
	\texttt{alejandro.rodriguez.collado@uva.es} \\
	\And
	\href{https://orcid.org/0000-0003-0254-4928}{\includegraphics[scale=0.06]{orcid.pdf}\hspace{1mm} Yolanda Larriba} \\
	Department of Statistics and Operations Research\\
 	Universidad of Valladolid, Valladolid, Spain \\
	\texttt{yolanda.larriba@uva.es} \\
	\And
	Adri\'an Lamela \\
	Department of Statistics and Operations Research\\
 	Universidad of Valladolid, Valladolid, Spain \\
	\texttt{adrianlamela@gmail.com} \\
	\And
	Christian Canedo \\
	Department of Statistics and Operations Research\\
 	Universidad of Valladolid, Valladolid, Spain \\
	\texttt{christian.canedo@alumnos.uva.es} \\
	\And
	\href{https://orcid.org/0000-0001-9638-8991}{\includegraphics[scale=0.06]{orcid.pdf}\hspace{1mm} Cristina Rueda} \\
	Department of Statistics and Operations Research\\
 	Universidad of Valladolid, Valladolid, Spain \\
	\texttt{cristina.rueda@uva.es} \\
}

\renewcommand{\shorttitle}{\textit{FMM}: Frequency Modulated M\"obius Models in R}

\begin{document}
\maketitle

\begin{abstract}
This paper is dedicated to the R package \textbf{FMM} which implements a novel approach to describe rhythmic patterns in oscillatory signals. The frequency modulated M\"obius (FMM) model is defined as a parametric signal plus a gaussian noise, where the signal can be described as a single or a sum of waves. The FMM approach is flexible enough to describe a great variety of rhythmic patterns. The \textbf{FMM} package includes all required functions to fit and explore single and multi-wave FMM models, as well as a restricted version that allows equality constraints between parameters representing a priori knowledge about the shape to be included. Moreover, the \textbf{FMM} package can generate synthetic data and visualize the results of the fitting process. The potential of this methodology is illustrated with examples of such biological oscillations as the circadian rhythm in gene expression, the electrical activity of the heartbeat and neuronal activity.
\end{abstract}


\section{Introduction} \label{sec:intro}
Oscillations naturally occur in a multitude of physical, chemical, biological, and even economic and social processes. Periodic signals appear, for example, during the cell-cycle, in biological time-keeping processes, in human heartbeats, in neuronal signals, in light emissions from certain types of stars, or in business cycles in economics, among many others. Three features typically describe the periodic nature of the oscillatory motion: period, amplitude and phase. The period is the time required for one complete oscillation. Within a period, a sum of monocomponent models characterized by the phase and amplitude parameters can be used to describe the rhythmic pattern of a signal \citep{Boashash:2016}. By varying the number of monocomponents and considering phase and amplitude parameters as fixed or variable, a large number of rhythmic signal representations can be found.

One of the most popular representations of oscillating signals is the Fourier decomposition (FD): a multicomponent representation with a fixed amplitude parameter. Its monocomponent version, the cosinor model (COS) \citep{Cornelissen:2014}, is widely used, in particular in chronobiology, with acceptable results when a sinusoidal shape response within a period is expected. Due to its widespread use, many software utilities are available. Particularly in R \citep{R}, the estimation of a COS model can be performed using \textbf{cosinor} \citep{Sachs:2014} and \textbf{cosinor2} packages \citep{Mutak:2018}. In addition, other packages from widely differing areas of knowledge have specific functions for fitting COS models. Such is the case of, for example, the function \textit{CATCosinor} in the \textbf{CATkit} package \citep{Gierke+Helget+Cronelissen:2018}, which implements tools for periodicity analysis; the function \textit{cosinor} in the \textbf{psych} package \citep{Revelle:2018}, dedicated to personality and psychological research; or the function \textit{cosinor} contained in a very recent package, \textbf{card} \citep{cardPakage}, which is dedicated to the assessment of the regulation of cardiovascular physiology. Recently, it has also been implemented in other languages such as \textbf{CosinorPy}, a cosinor python package \citep{Moskon:2020}. The COS model is easy to use and interpret with symmetrical patterns. However, asymmetric shapes are not captured properly by COS. When the waveform is non-sinusoidal, the use of multiple components analysis to fit a model consisting of a sum of several periodical functions is recommended. However, the multicomponent FD models, developed to provide flexibility from COS, often require the use of a large number of components resulting in serious over-fitting issues. 

In recent years, alternative methods, mostly nonparametric statistical methods, have been developed and used for analyzing rhythmicity, especially in biological data sets. Some very popular ones, such as the JTK\_Cycle \citep{Hughes+Hogenesch+Kornacker:2010}, wrongly assume that any underlying rhythms have symmetric waveforms. Others, such as RAIN \citep{Thaben+Westermark:2014}, designed to detect more diverse wave shapes including asymmetric patterns, are not focused on modeling but on detecting rhythmic behavior in sets of data. Thus, they are not useful to describe the underlying oscillatory phenomena. The proliferation of methodology in this field has been accompanied by software developments. This is the case, for example, of the \textbf{DiscoRhythm} R package \citep{Calucci+Kris+Li+al:2020}, very recently available on Bioconductor (\url{https://bioconductor.org/packages/DiscoRhythm}) with a web interface based on the R \textbf{Shiny} platform \citep{Chang+Cheng+Allaire+Xie+McPherson:2020}. This tool allows four popular approaches to be used, including the COS model and JTK\_Cycle, to discover biological rhythmicity. Another recent example is the \textbf{CircaCompare} \citep{Parsons+Garner+Oster+Rawashdeh:2019}, an R package implemented for modeling cosinusoidal curves by nonlinear regression and available from GitHub (\url{https://github.com/RWParsons/circacompare}). Also hosted on GitHub, we can find the \textbf{LimoRhyde} R package (\url{https://github.com/hugheylab/limorhyde}) for the differential analysis of rhythmic transcriptome data, based on fitting linear models \citep{Singer+Hughey:2019}. 
      
Motivated by the need for a flexible, interpretable and parametric methodology to fit rhythmic patterns, our research group recently proposed the frequency modulated M\"obius (FMM) model \citep{Rueda+Larriba+Peddada:2019}. The FMM is an additive nonlinear parametric regression model capable of adapting to non-sinusoidal shapes and whose parameters are easily interpretable. The single component model has been shown to successfully fit data as diverse as circadian clock signals, hormonal levels data or light data from distant stars. In addition, for more complex oscillatory signals, a multicomponent model of order $m$, denoted as FMM$_m$, which includes $m$ single FMM components, can be used. This is, for example, the case for describing electrocardiography (ECG) signals. The FMM$_{ecg}$ signal, presented in \cite{Rueda+Larriba+Lamela:2021}, is defined as the combination of five single FMM components. Another interesting area where the FMM approach has already shown its usefulness is in electrophysiological neuroscience. Specifically, we have proposed FMM methodology for modeling neuronal action potential (AP) curves, oscillating signals that measure the difference between the electrical potential inside and outside the cell \citep[see][]{Rueda+Rodriguez+Larriba:2021, RodriguezCollado+Rueda:2021}. An FMM$_2$ model provides an accurate fitting for a single AP curve; whereas series of AP curves with similar repetitive spikes can be efficiently fitted by the FMM$_{ST}$ model, a restricted version of the multicomponent FMM model.

In this work we introduce the \textbf{FMM} package, programmed in R and available from the Comprehensive R Archive Network (CRAN) at \url{http://CRAN.R-project.org/package=FMM}. The package implements all required functions to fit and explore single and multicomponent FMM models, as well as a restricted multicomponent version. In addition, the \textbf{FMM} package provides functions to generate synthetic data and visualize the results of the fitted model. Furthermore, its use is illustrated in the aforementioned applications. The remainder of this paper is organized as follows: Section~\ref{sec:models} provides a brief overview of both mono and multicomponent FMM models (\ref{subsec:monoFMM} and \ref{subsec:multiFMM}, respectively). The FMM$_m$ model with equality constraints is presented in Section~\ref{subsec:restFMM}. Section~\ref{sec:implementation} is dedicated to the implementation details of the \textbf{FMM} package. Through simulated examples, Section~\ref{sec:use} introduces the basic usage of the package, including the data generation and the fitting, as well as the visualization of the results. In Section~\ref{sec:realExamp}, the \textbf{FMM} package performance is shown through three application areas governed by oscillatory systems: chronobiology, ECG and neuroscience. Finally, a summary is provided in Section~\ref{sec:summary}.

\section{Frequency modulated M\"obius (FMM) model} \label{sec:models}
All the methodological details that justify the mathematical formulation of the FMM models are given in \citealt{Rueda+Larriba+Peddada:2019}.

At the time point $t$, a single FMM wave is defined as $ W(t; \upsilon) = A \cos(\phi(t; \alpha, \beta, \omega))$ where $\upsilon = (A, \alpha, \beta, \omega)^{\prime} $, $A \in \Re^{+}$ represents the wave amplitude and, 
\begin{equation} \label{eq:phase}
\phi(t; \alpha, \beta, \omega )= \beta + 2\arctan(\omega \tan(\frac{t - \alpha}{2}))
\end{equation}
the wave phase. The phase angle $\phi$ of an FMM wave is defined using the M\"obius link \citep[see][]{Downs+Mardia:2002, Kato+Shimizu+Shieh:2008} rather than the linear link function as in the COS model. The M\"obius link provides much more flexibility to describe non-sinusoidal patterns. 
Without loss of generality, we assume that the time point $t \in [0, 2\pi]$. Otherwise it can be transformed into $t^{\prime} \in [t_0, T + t_0]$ by $t = \frac{(t^{\prime} - t_0)2\pi}{T}$.

Each of the four parameters of an FMM wave characterizes some aspect of a rhythmic pattern. $A$ describes the amplitude of the signal, while $\alpha$, $\beta$ and $\omega$ describe the wave phase. $\alpha \in [0, 2\pi]$ is a translation parameter and a wave location parameter in the real space, whereas $\beta \in [0, 2\pi]$ and $\omega \in [0, 1]$ describe the wave shape. To be precise, assuming $\alpha = 0$, the unimodal symmetric waves are characterized by values of $\beta$ close to $0$, $\pi$ or $2\pi$. When $\beta = \frac{\pi}{2}$ or $\beta = \frac{3\pi}{2}$, extreme asymmetric patterns are described. Moreover, a value of $\omega$ close to zero describes an extreme spiked wave and, as $\omega$ value increases, the pattern is increasingly smoother. When $\omega = 1$, a sinusoidal wave is described and the FMM model matches the COS model where $\varphi = \beta - \alpha$ is the acrophase parameter. 

Two important features of a wave are the peak and trough, defined as the highest and lowest points above and below the rest position, respectively. In many applications, the peak and trough times  could be very useful tools to extract practical information of a wave, since they capture  important aspects of the dynamics. These two interesting parameters can be directly derived from the main parameters of an FMM wave as,
\begin{eqnarray} \label{eq:monoFMM:fiducial}
t^U & = & \alpha + 2\arctan(\frac{1}{\omega}\tan(-\frac{\beta}{2})) \\
t^L & = & \alpha + 2\arctan(\frac{1}{\omega}\tan(\frac{\pi-\beta}{2})) \nonumber
\end{eqnarray}
where $t^U$ and $t^L$ denote the peak and trough times, respectively. 

\subsection{Monocomponent FMM model} \label{subsec:monoFMM}
Let $X(t_i)$, $t_1 < t_2 < \dots < t_n$ be the vector of observations.
The monocomponent FMM model is defined as follows:
\begin{equation} \label{eq:monoFMM}
X(t_i) = M + W(t_i; \upsilon) + e(t_i), \quad i = 1,\dots,n 
\end{equation} 
where $M \in \Re$ is an intercept parameter describing the baseline level of the signal, $W(t_i; \upsilon)$ is an FMM wave, and it is assumed that the errors $e(t_i)$ are independent and normally distributed with zero mean and a common variance $\sigma^2$.

\subsubsection{Estimation algorithm} \label{subsubsec:monoFMM:estimation}
A two-step algorithm to estimate monocomponent FMM model parameters is proposed. We now describe the substantial details of each stage of the algorithm.    

\textit{\textbf{Step 1:} Initial parameter estimation}. 
A two-way grid search over the choice of $(\alpha , \omega)$ parameters is performed. For each pair of $(\alpha , \omega)$ fixed values, the estimates for $M$, $A$ and $\beta$ are obtained by solving a least square problem as detailed below. 

The model for a single FMM component can be written as:
\begin{equation} \label{eq:monoFMM:muiParam}
X(t_i) = M + A\cos(t_{i}^{*} + \varphi)+ e(t_i)  
\end{equation}
where $t_{i}^{*} = \alpha + 2\arctan(\omega \tan(\frac{t_i - \alpha}{2}))$,  $\varphi = \beta - \alpha$, and $e(t_i) \sim N(0, \sigma^2)$ for $i = 1, \dots, n$.

Using trigonometric angle sum identity, the model can be rewritten as:
\begin{equation} \label{eq:monoFMM:reWrite}
X(t_i) = M + \delta z_i + \gamma w_i + e(t_i) 
\end{equation}
where $\delta = A \cos(\varphi)$, $\gamma = -A \sin(\varphi)$, $z_i = \cos(t_i^*)$ and $w_i = \sin(t_i^*)$.

Since $\alpha$ and $\omega$ are fixed, the estimates for $M$, $\delta$ and $\gamma$ are obtained by minimizing the residual sum of the squares (RSS),
\begin{equation} \label{eq:monoFMM:SSR}
RSS = \sum_{i=1}^{n}(X(t_i)-(\hat{M} + \hat{\delta} z_i + \hat{\gamma} w_i))^2 
\end{equation}
And the estimates for $M$, $A$ and $\beta$ are straightforward to derive as follows,
\begin{eqnarray} \label{eq:monoFMM:estimates}
  \hat{M} & = & \bar{X} - \hat{\delta} \sum_{i=1}^{n} z_i - \hat{\gamma} \sum_{i=1}^{n} w_i\\
  \hat{A} & = & \sqrt{\hat{\delta}^2 + \hat{\gamma}^2}  \\
  \hat{\beta} & = & \alpha + \varphi  
\end{eqnarray}
The best combination of $(\alpha, \omega)$ values, with the lowest RSS, is retained and the corresponding estimates are the initial parameter estimation values.

\textit{\textbf{Step 2:} Optimization}. In the second step, the Nelder-Mead optimization method \citep{Nelder+Mead:1965} is used to obtain the final FMM parameter estimates. Nelder-Mead is a heuristic search method commonly used to minimize or maximize an objective function in a multidimensional space. 

\subsection{Multicomponent FMM model} \label{subsec:multiFMM}
A multicomponent FMM model of order $m$, denoted by FMM$_m$, is defined as
\begin{eqnarray} \label{eq:multiFMM}
X(t_i) = M + \sum_{J=1}^{m}W(t_i; \upsilon_J) + e(t_i) \\
t_1 < t_2 < \dots < t_n; i = 1, \dots, n  \nonumber
\end{eqnarray} 
where $W(t_i; \upsilon_J)$, hereinafter denoted by $W_J(t_i)$, is the Jth FMM wave and,
\begin{itemize}
  \item $M \in \Re$
  \item $ \upsilon_J = (A_J, \alpha_J, \beta_J, \omega_J)^{\prime} \in  \Re^{+} \times [0,2\pi] \times [0,2\pi] \times [0,1]$; $J = 1, \dots ,m$ 
  \item $ \alpha_1 \leq \alpha_2 \leq \dots \leq \alpha_m \leq \alpha_1$
  \item $(e(t_1), \dots , e(t_n))^{\prime} \sim N_n(0, \sigma^2I_n) $
\end{itemize}

\subsubsection{Model adequacy} \label{subsubsec:multiFMM:R2}
The goodness of fit of an FMM model is measured with the $R^2$ statistic that represents the proportion of the variance explained by the model out of the total variance, that is:
\begin{equation} \label{eq:multiFMM:R2}
R^2 = 1 - \frac{\sum_{i=1}^{n} (X(t_i) - \hat{X}(t_i))^2}{\sum_{i=1}^{n} (X(t_i) - \bar{X})^2}
\end{equation}
where $\hat{X}(t_i)$ represents the fitted value at $t_i, i=1, \dots ,n$.

\subsubsection{Estimation algorithm} \label{subsubsec:multiFMM:estimation}
An iterative backfitting algorithm is proposed to derive estimates for the FMM parameters. Let $\hat{W}_J^{(k)}(t_i)$ denote the fitted values from the Jth FMM wave at $t_i, i=1,...,n$ in the kth iteration. The algorithm is structured as follows:
\begin{enumerate}
   \item [1] Initialize. Set $\hat{W}_1^{(0)}(t_i) = \dots = \hat{W}_m^{(0)}(t_i) =
    0$. 	
   \item [2] Backfitting step. For $J = 1, \dots ,m$, calculate
      \begin{equation} \label{eq:multiFMM:partialres}
        r_J^{(k)}(t_i) = X(t_i) - \sum_{I = 1}^{J-1} \hat{W}_I^{(k)}(t_i) - 
        \sum_{I = J+1}^{m} \hat{W}_I^{(k-1)}(t_i)
      \end{equation}
     and fit a monocomponent FMM model to $r_J^{(k)}(t_i)$ obtaining 
     $\hat{\alpha}_J^{(k)}$, $\hat{\beta}_J^{(k)}$, $\hat{\omega}_J^{(k)}$ and 
     $\hat{W}_J^{(k)}(t_i)$.
    \item [3] Repeat the backfitting step until the stopping criterion is reached. 
    The stopping criterion is defined as the difference between the explained
    variability in two consecutive iterations: $R_k^2 - R_{k-1}^2 \leq C$, where 
    $R_k^2$ (defined in Equation~\ref{eq:multiFMM:R2}) is the proportion of variance
    explained by the model in the kth iteration and $C$ a constant.  
   \item [4] $\hat{M}$ and $\hat{A_J}$ are derived by solving
   \begin{equation} \label{eq:multiFMM:backfitting}
    \min_{M \in \Re; A_J \in \Re^+} \sum_{i=1}^n (
    X(t_i) -  M - \sum_{J=1}^{m} A_{J}\cos(\hat{\phi}_J(t_i)))^2 
   \end{equation} 
   where $\hat{\phi}_J(t_i) = \phi(t_i; \hat{\alpha}_J, \hat{\beta}_J, \hat{\omega}_J)$ defined in Equation~\ref{eq:phase}.
\end{enumerate}

\subsection{Restricted multicomponent FMM model}\label{subsec:restFMM}
Modeling signals with repetitive shape-similar waves can be very useful in some applications \citep[see][]{RodriguezCollado+Rueda:2021}. In order to obtain more efficient estimators, equality constraints are imposed on the $\beta$ and $\omega$ parameters of an FMM$_m$ model. In particular, we add $d$ blocks of restrictions on $\beta$ parameters:
\begin{eqnarray} \label{eq:multiFMM:betaRestr}
  &\beta_1 = \dots = \beta_{m_1} \\
  &\beta_{m_1+1} = \dots = \beta_{m_2} \nonumber \\
  &\dots \nonumber \\
  &\beta_{m_{d-1}+1} = \dots = \beta_{m_d} \nonumber
\end{eqnarray}
and/or $d^{\prime}$ blocks of restrictions on $\omega$ parameters:
\begin{eqnarray} \label{eq:multiFMM:omegaRestr}
  &\omega_1 = \dots = \omega_{m_1^{\prime}} \\
  &\omega_{m_1^{\prime}+1} = \dots = \omega_{m^{\prime}_2} \nonumber \\
  &\dots \nonumber \\
  &\omega_{m_{d^{\prime}-1}^{\prime}+1} = \dots = \omega_{m^{\prime}_{d^{\prime}}} \nonumber
\end{eqnarray}

The parameter estimation problem is solved by an adaptation of the standard procedure.

\subsubsection{FMM$_m$ estimation algorithm with restrictions on the $\beta$ parameters}\label{subsubsec:restFMM:beta}
Given the unrestricted estimates obtained in step $3$, the estimates for $\beta_1, \beta_{m_1+1}, \dots, \beta_{m_{d-1}+1}$ under equality restrictions (Equation~\ref{eq:multiFMM:betaRestr}) are computed as follows:

\begin{eqnarray*}
&\hat{\beta}_J^* = \mathrm{angularMean}(\hat{\beta}_1, \dots ,\hat{\beta}_{m_1})& 
J = 1, \dots ,m_1 \\
&\hat{\beta}_J^* = \mathrm{angularMean}(\hat{\beta}_{m_1+1}, \dots ,\hat{\beta}_{m_2})& 
J = m_1 + 1, \dots ,m_2 \\
&\dots \\
&\hat{\beta}_J^* = \mathrm{angularMean}(\hat{\beta}_{m_{d-1}+1}, \dots ,\hat{\beta}_{m_d})&
J = m_{d-1} + 1, \dots ,m_d
\end{eqnarray*}
  
Then, the algorithm continues to the next step.

\subsubsection{FMM$_m$ estimation algorithm with restrictions on the $\omega$ parameters}\label{subsubsec:restFMM:omega}
A block is defined as the set of FMM waves with common $\omega$ parameter. When constraints for the $\omega$ parameters are incorporated, the grid search for the different $\omega$ values is outside the backfitting loops. When the number of blocks is large, the estimation procedure can be computationally unaffordable. 
In order to reduce the execution time, a two-nested backfitting algorithm is proposed. 
In the outer backfitting loop, a block is fitted. In the inner loop, the FMM waves belonging to the same block are estimated. This procedure generates a close to optimal solution and is a less computationally expensive alternative. 

\section{\textbf{FMM} package: Implementation details} \label{sec:implementation}
The \textbf{FMM} code makes use of the \textbf{doParallel} package \citep{Weston:2020} to embed parallelization for the restricted version of the fitting process. Several utilities from the \textbf{ggplot2} \citep{Wickham:2016} and \textbf{RColorBrewer} \citep{Neuwirth:2014} packages are occasionally necessary for the visualization of the fitted models.  

The implementation of \textbf{FMM} is divided into four main functionalities described from Sections~\ref{subsec:impl:FMMModel} to \ref{subsec:impl:FMMSimulate}: the fitting of the FMM models, the new $S4$ object of class `FMM', the graphical visualization of the fittings and the simulation of synthetic data.

Some general details about the functions contained in the \textbf{FMM} package are shown in Table~\ref{tab:functions}. 

\begin{table}[!htbp] 
\centering
\begin{tabular}{p{7cm}p{8.5cm}}
\hline
\textbf{Function}                 &  \textbf{Description}\\ \hline
\textbf{\textit{Fitting function}}\\
$\mathrm{fitFMM(vData, timePoints, nback, ...)}$  & Estimates an FMM$_{\mathrm{nback}}$
													model to $\mathrm{vData}$ observed
													at $\mathrm{timePoints}$. \\ \hline
\textbf{\textit{Utility functions}}\\
$\mathrm{plotFMM(objFMM, ...)}$          		  & Graphically displays an object of 
												    class `FMM'.\\
$\mathrm{generateFMM(M, A, alpha, beta,
 omega, ...)}$  								  &	Simulates values from an FMM model.\\
$\mathrm{getFMMPeaks(objFMM, ...)}$ 			  & Estimates peak and trough times,
													together with signal values at those
													times, for each FMM wave.\\
$\mathrm{extractWaves(objFMM)}$					  & Extracts individual contribution to
												    the fitted values of each FMM wave.\\
\hline
\textbf{\textit{Standard methods for objects of class `FMM'}}\\
\multicolumn{2}{l}{$\mathrm{summary()}$, $\mathrm{show()}$, $\mathrm{coef()}$, $\mathrm{fitted()}$ }\\
\hline 
\end{tabular}
\caption{\label{tab:functions} Summary of the fitting, utility functions and standard methods implemented in \textbf{FMM} package.}
\end{table}

\subsection{Fitting an FMM model}\label{subsec:impl:FMMModel}
An FMM model can be fitted using the main function \textit{fitFMM()}. The description and default values of its inputs arguments are shown in Table~\ref{tab:argumentsFit}. 

The fitting function \textit{fitFMM()} requires the $\mathrm{vData}$ input argument, which contains the data to be fitted. Two other arguments can be used to control a basic fitting: $\mathrm{timePoints}$, which contains the specific time points of the single period; and $\mathrm{nback}$, with the number of FMM components to be fitted. For some applications, such as the study of circadian rhythms, data are collected over multiple periods. This information is received by the \textit{fitFMM()} function through the input argument $\mathrm{nPeriods}$. When $\mathrm{nPeriods} > 1$, the FMM fitting is carried out by averaging the data collected at each time point across all considered periods.

\begin{table}[!htbp] 
\centering
\begin{tabular}{p{3cm}p{2.5cm}p{9.5cm}}
\hline
\textbf{Argument}     &  \textbf{Default value}	 &  \textbf{Description}\\ \hline
$\mathrm{vData}$      &  no default value 		 &  Data to be fitted by an FMM model. \\
$\mathrm{nPeriods}$   &  $1$                     &  Number of periods at which 
                                                    $\mathrm{vData}$ is observed.\\
$\mathrm{timePoints}$ &  $\mathrm{NULL}$         &  Vector of time points per period 
                                                    at which data is observed. When
                                                    $\mathrm{timePoints = NULL}$ an
                                                    equally spaced sequence from $0$ to
												    $2\pi$ will be assigned. \\
$\mathrm{nback}$      &  $1$                     &  Number of FMM components to be
                                                    fitted.\\
$\mathrm{beta
Restrictions}$        &  $1:\mathrm{nback}$      &  $\beta$ constraint vector.\\
$\mathrm{omega
Restrictions}$        &  $1:\mathrm{nback}$      &  $\omega$ constraint vector.\\
$\mathrm{maxiter}$    &  $\mathrm{nback}$        &  Maximum number of iterations for the
                                                    backfitting algorithm.\\
$\mathrm{stop
Function}$            &  $\mathrm{alwaysFalse}$  &  Function to check the criterion
                                                    convergence for the backfitting
                                                    algorithm.\\
$\mathrm{lengthAlpha
Grid}$                &  $48$                   &  Grid resolution of the parameter 
                                                   $\alpha$.\\
$\mathrm{lengthOmega
Grid}$                &  $24$                   &  Grid resolution of the parameter 
                                                   $\omega$.\\
$\mathrm{numReps}$    &  $3$                    &  Number of times $(\alpha, \omega)$
												   parameters are refined.\\       
$\mathrm{show
Progress}$            &  $\mathrm{TRUE}$        &  $\mathrm{TRUE}$ to display a progress
                                                   indicator on the console.\\                        
$\mathrm{showTime}$   &  $\mathrm{TRUE}$        &  $\mathrm{TRUE}$ to display execution
                                                   time on the console.\\                                
$\mathrm{parallelize}$&  $\mathrm{FALSE}$       &  $\mathrm{TRUE}$ to use parallelized
                                                   procedure to fit restricted 
                                                   FMM$_m$ model. \\
\hline 
\end{tabular}
\caption{\label{tab:argumentsFit} Description of the input arguments of the \textit{fitFMM()} function and their default values.}
\end{table}

There are three key issues in the fitting process: the grid search of the pair $(\alpha, \omega)$ to solve the estimation problem of a single FMM wave, the backfitting algorithm used for the estimation of the multicomponent models, and the incorporation of restrictions on $\beta$ and $\omega$ parameters. Each of these issues is controlled by several arguments described below.

\begin{itemize}
\item \textbf{Grid search of the pair $(\alpha, \omega)$}. The $\mathrm{lengthAlphaGrid}$ and $\mathrm{lengthOmegaGrid}$ arguments are used to set the grid resolution by specifying the number of equally spaced $\alpha$ and $\omega$ values. Thus, the objetive function will be evaluated a total number of $(\mathrm{lengthAlphaGrid})\times(\mathrm{lengthOmegaGrid})$ times, so when both arguments are large, the computational demand can be high. By reducing the size of the sequences of the $\alpha$ and $\omega$ parameters, the algorithm will be computationally more efficient. However, it may fail to obtain an accurate estimation if the grid resolution is too sparse. 
An implemented option to fine-tune the estimation of the parameters is to repeat the fitting process a $\mathrm{numReps}$ of times, in such a way that, at each repetition, a new two-dimensional grid of $(\alpha, \omega)$ points is created around the previous estimates.

\item \textbf{Backfitting algorithm}. The argument $\mathrm{maxiter}$ sets the maximum number of backfitting iterations. Through the argument $\mathrm{stopFunction}$, it is possible to set a stopping criterion. Two criteria have been implemented as stop functions in this package. When $\mathrm{stopFunction = alwaysFalse}$, $\mathrm{maxiter}$ iterations will be forced. If $\mathrm{stopFunction = R2()}$, the algorithm will be stopped when the difference between the explained variability in two consecutive iterations is less than a value pre-specified in the $\mathrm{difMax}$ argument. 

\item \textbf{Restrictions}. The arguments $\mathrm{betaRestrictions}$ and $\mathrm{omegaRestrictions}$ set the equality constraints for the $\beta$ and $\omega$ parameters, respectively. For the unrestricted case, $\mathrm{betaRestrictions} = 1:\mathrm{nback}$ and $\mathrm{omegaRestrictions} = 1:\mathrm{nback}$. To add restrictions, `integer' vectors of length $m$ can be passed to these arguments, so that positions with the same numeric value correspond to FMM waves whose parameters, $\beta$ and/or $\omega$, are forced to be equal. Since restricted fitting can be computationally intensive, especially for constraints on $\omega$ parameters, a parallel processing implementation can be used when the argument $\mathrm{parallelize = TRUE}$. 
\end{itemize}

\subsection{Object of class `FMM'}\label{subsec:impl:FMMClass}
The \textit{fitFMM()} function outputs an $S4$ object of class `FMM' which contains the slots presented in Table~\ref{tab:slotsFMM}. 

\begin{table}[!ht] 
\centering
\begin{tabular}{p{3cm}p{12cm}}\hline
   \textbf{Slot}             &  \textbf{Description} \\ \hline
   $\mathrm{timePoints}$     &  A `numeric' vector containing the time points for each 
                                data point if one single period is observed.\\
   $\mathrm{data}$           &  A `numeric' vector containing the data to be fitted to 
                                an FMM model. Data could be collected over multiple
                                periods.\\
   $\mathrm{summarizedData}$ &  A `numeric' vector containing the summarized data at 
                                each time point across all considered periods.\\                        
   $\mathrm{nPeriods}$       &  A `numeric' value containing the number of periods in
                                $\mathrm{data}$.\\
   $\mathrm{fittedValues}$   &  A `numeric' vector of the fitted values by the FMM 
                                model.\\
   $\mathrm{M}$              &  A `numeric' value of the estimated intercept parameter 
                                $M$.\\ 
   $\mathrm{A}$              &  A $m$-element `numeric' vector of the estimated FMM wave
                                amplitude parameter(s) $A$.\\        
   $\mathrm{alpha}$          &  A $m$-element `numeric' vector of the estimated FMM wave
                                phase translation parameter(s) $\alpha$.\\  
   $\mathrm{beta}$           &  A $m$-element `numeric' vector of the estimated FMM wave
                                skewness parameter(s) $\beta$.\\ 
   $\mathrm{omega}$          &  A $m$-element `numeric' vector of the estimated FMM wave
                                kurtosis parameter(s) $\omega$.\\                                  
   $\mathrm{SSE}$            &  A `numeric' value of the residual sum of squares 
                                values.\\
   $\mathrm{R2}$             &  A $m$-element `numeric' vector specifying the explained
                                variance by each of the fitted FMM components.\\  
   $\mathrm{nIter}$          &  A `numeric' value containing the number of iterations
   							    of the backfitting algorithm.\\  
   \hline                                          
\end{tabular}
\caption{\label{tab:slotsFMM} Summary of the slots of the $S4$ object of class `FMM' resulting from fitting an FMM model with $m$ components.}
\end{table}

The standard methods implemented for the class `FMM' include the functions \textit{summary()}, \textit{show()}, \textit{coef()} and \textit{fitted()}. These methods display relevant information of the FMM fitting, and provide the estimated parameters and fitted values. In addition, two more specific functions have been implemented. 
Through the \textit{extractWaves()} function, the individual contribution of each FMM wave to the fitted values can be extracted. Finally, the location of the peak and trough of each FMM wave, as well as the value of the signal at these time points, can be estimated using the \textit{getFMMPeaks()} function. The required argument of all these methods and functions is an object of the class `FMM'. Particularly, \textit{getFMMPeaks()} has an optional argument: $\mathrm{timePointsIn2pi}$, that forces the peak and trough locations to be returned into the interval from $0$ to $2\pi$ when it is $\mathrm{TRUE}$.

\subsection{Plotting FMM models}\label{subsec:impl:FMMPlot}
The \textbf{FMM} package includes the function \textit{plotFMM()} to visualize the results of an FMM fit. The arguments of this function are summarized in Table~\ref{tab:argumentsPlot}.

\begin{table}[!htbp] 
\centering
\begin{tabular}{p{4.5cm}p{3cm}p{7.5cm}}
\hline
  \textbf{Argument}    &  \textbf{Default value}  &  \textbf{Description}\\ \hline
  $\mathrm{objFMM}$    &  no default value        & The object of class `FMM' to be
                                                    plotted. \\
  $\mathrm{components}$&  $\mathrm{FALSE}$        & $\mathrm{TRUE}$ to display a plot
                                                    of components.\\
  $\mathrm{
  plotAlongPeriods}$   &  $\mathrm{FALSE}$        & $\mathrm{TRUE}$ to plot more than 1
                                                    period. \\
  $\mathrm{
  use\_ggplot2}$       &  $\mathrm{FALSE}$        & $\mathrm{TRUE}$ to plot with
													 \textbf{ggplot2} package.\\
  $\mathrm{legendIn
  ComponentsPlot}$     &  $\mathrm{TRUE}$         & $\mathrm{TRUE}$ to indicate if 
                                                     a legend should be plotted in 
                                                     the component plot.\\
  $\mathrm{textExtra}$ & empty string             &  Extra text to be added to the 
                                                     title of the plot.\\
\hline 
\end{tabular}
\caption{\label{tab:argumentsPlot} Description of the input arguments of the \textit{plotFMM()} function and their default values.}
\end{table}

An object of class `FMM' can be plotted in two ways (see Figure~\ref{f:sim1}). The default graphical representation will be a plot on which original data (as points) and the fitted model (as a line) are plotted together (left panel in Figure~\ref{f:sim1}). The other possible representation is a component plot for displaying each centered FMM wave separately (right panel in Figure~\ref{f:sim1}). Set the boolean argument $\mathrm{components = TRUE}$ to show a component plot. When $\mathrm{legendInComponentsPlot = TRUE}$, a legend appears at the bottom of the component plot to indicate the represented waves. The argument $\mathrm{textExtra}$ allows an extra text to be added to the title of both graphical representations.

As mentioned above, in some cases, data are collected from different periods. All periods can be displayed simultaneously on the default plot using $\mathrm{plotAlongPeriods = TRUE}$. For the component plot, this argument is ignored.

The argument $\mathrm{use\_ggplot2}$ provides a choice between building the plot using base R \textbf{graphics} or \textbf{ggplot2} packages. By default, the \textbf{graphics} package is used. When $\mathrm{use\_ggplot2 = TRUE}$, a more aesthetic and customizable plot is created using the \textbf{ggplot2} package. 

\subsection{Simulating data from an FMM model}\label{subsec:impl:FMMSimulate}
Data from an FMM model can be easily simulated using the function \textit{generateFMM()} of the package \textbf{FMM}. All input arguments of this function are shown in Table~\ref{tab:argumentsGenerate}, along with a short description and their default values. 

\begin{table}[!h] 
\centering
\begin{tabular}{p{2cm}p{5cm}p{8cm}}
\hline
  \textbf{Argument}	   &  \textbf{Default value}   &  \textbf{Description}\\ \hline
  $\mathrm{M}$         &  no default value         &  Value of the intercept parameter 
                                                      $M$. \\
  $\mathrm{A}$         &  no default value         &  Vector of the FMM wave amplitude
                                                      parameter $A$. \\
  $\mathrm{alpha}$     &  no default value         &  Vector of the FMM wave phase
                                                      translation parameter $\alpha$. \\
  $\mathrm{beta}$      &  no default value         &  Vector of the FMM wave skewness
                                                      parameter $\beta$. \\
  $\mathrm{omega}$     &  no default value         &  Vector of the FMM wave kurtosis
                                                      parameter $\omega$. \\
  $\mathrm{from}$      &  $0$                      &  Initial time point of the simulated
                                                      data.\\
  $\mathrm{to}$        &  $2\pi$                   &  Final time point of the simulated
                                                      data.\\
  $\mathrm{length.out}$&  $100$                    &  Desired length of the simulation.\\
  $\mathrm{timePoints}$&  $\mathrm{seq(from, 
                          to, length = 
                          length.out)}$            &  Time points at which the data will
                                                      be simulated.\\
  $\mathrm{plot}$      & $\mathrm{TRUE}$           &  $\mathrm{TRUE}$ when the simulated
                                                      data should be drawn on a plot.\\
  $\mathrm{outvalues}$ & $\mathrm{TRUE}$           &  $\mathrm{TRUE}$ when the numerical
                                                      simulation should be returned.\\
  $\mathrm{sigmaNoise}$& $0$                       &  Standard deviation of the gaussian
                                                      noise to be added.\\
\hline 
\end{tabular}
\caption{\label{tab:argumentsGenerate} Description of the input arguments of the  \textit{generateFMM()} function and their default values.}
\end{table}

The main arguments of this function are $\mathrm{M}$, $\mathrm{A}$, $\mathrm{alpha}$, $\mathrm{beta}$ and $\mathrm{omega}$, whereby the values of the FMM model parameters are passed to the function. All these arguments are `numeric' vectors of length $m$, except $\mathrm{M}$, which has length $1$. Longer and smaller vectors will be truncated or replicated as appropriate. 

By default, the data will be simulated at a sequence of $100$ equally spaced time points from $0$ to $2\pi$. The arguments $\mathrm{from}$, $\mathrm{to}$ and $\mathrm{length.out}$ control such sequences. The sequence can also be manually set using the argument $\mathrm{timePoints}$, in which case $\mathrm{from}$, $\mathrm{to}$ and $\mathrm{length.out}$ will be ignored.

The user can add a gaussian noise by argument $\mathrm{sigmaNoise}$. A positive `numeric' value sets the corresponding standard deviation of the gaussian noise to be added. To create the normally distributed noise, the \textit{rnorm()} function is used.

The arguments $\mathrm{plot}$ and $\mathrm{outvalues}$, both boolean values, determine the output of the \textit{generateFMM()} function. When $\mathrm{outvalues = TRUE}$, a `list' with input parameters, time points and simulated data is returned. These elements are named $\mathrm{input}$, $\mathrm{t}$ and $\mathrm{y}$, respectively.  In addition, a scatter plot of $\mathrm{y}$ against $\mathrm{t}$ can be drawn by setting $\mathrm{plot = TRUE}$. 

\section{Basic usage of the \textbf{FMM} package} \label{sec:use}
The two examples bellow, based on FMM synthetic data, illustrate the basic uses and capabilities of the functions implemented in the \textbf{FMM} package. The first example is based on a simulated data set from a multicomponent FMM model and illustrates the fitting process and the presentation of its results. The second is devoted to the restricted version of FMM$_m$ models and serves to exemplify the control of the grid search related arguments to reduce the execution time. 

\subsection{Example 1: A multicomponent FMM data set}\label{subsec:use:multFMM}
The first simulated data set consists of $100$ observations from an FMM$_2$ model with intercept parameter $M = 0$ and two FMM waves defined by the parameters $ \upsilon_1 = (A_1 = 2, \alpha_1 = 1.5, \beta_1 = 0.2, \omega_1 = 0.1)^{\prime}$ and $ \upsilon_2 = (A_2 = 2, \alpha_2 = 3.4, \beta_2 = 2.3, \omega_2 = 0.2)^{\prime}$, respectively. The standard deviation of the error term is set at $\sigma = 0.3$. We use the function \textit{generateFMM()} to simulate this data set. A \textit{set.seed()} statement is used to guarantee the reproducibility of the results. 
\begin{lstlisting}
R> library("FMM")
R> set.seed(15)
R> fmm2.data <- generateFMM(M = 0, A = rep(2,1), 
                            alpha = c(1.5, 3.4), 
                            beta = c(0.2, 2.3), 
                            omega = c(0.1, 0.2),
                            plot = FALSE, outvalues = TRUE, sigmaNoise = 0.3) 
R> str(fmm2.data)

List of 3
 $ input:List of 5
  ..$ M    : num 0
  ..$ A    : num [1:2] 2 2
  ..$ alpha: num [1:2] 1.5 3.4
  ..$ beta : num [1:2] 0.2 2.3
  ..$ omega: num [1:2] 0.1 0.2
 $ t    : num [1:100] 0 0.0635 0.1269 0.1904 0.2539 ...
 $ y    : num [1:100] 1.26 2.19 2.08 3.03 3.44 ...
\end{lstlisting}

The estimation of an FMM$_2$ can be performed by setting $\mathrm{nback = 2}$ in the fitting function \textit{fitFMM()}.
\begin{lstlisting}
R> fit.fmm2 <- fitFMM(vData = fmm2.data$y, timePoints = fmm2.data$t, 
                      nback = 2 , showTime = FALSE)

|--------------------------------------------------|
|==================================================|
Stopped by reaching maximum iterations ( 2 iterations ) 
\end{lstlisting}
When the argument $\mathrm{showProgress = TRUE}$ (its default value), a progress indicator and information about the stopping criterion of the backfitting algorithm will be displayed on the console. 
By default, $\mathrm{maxiter}$ iterations will be forced, but we can use the argument $\mathrm{stopFunction}$ to modify the criteria. For example, to continue the search until there is an improvement, in terms of explained variability, of less than $1\%$, the \textit{R2()} function can be used specifying the $\mathrm{difMax = 0.01}$ argument. 
\begin{lstlisting}
R> fit.fmm2 <- fitFMM(vData = fmm2.data$y, timePoints = fmm2.data$t, nback = 2,
                      showTime = FALSE, maxiter = 5, stopFunction = R2(difMax = 0.01))

|--------------------------------------------------|
|==================================================|
Stopped by the stopFunction ( 4 iterations ) 
\end{lstlisting}

The results are displayed by the function \textit{summary()}:
\begin{lstlisting}
R> summary(fit.fmm2)

Title:
FMM model with 2 components

Coefficients:
M (Intercept): -0.2188
                  A  alpha   beta  omega
FMM wave 1:  1.9615 3.3743 2.2275 0.2053
FMM wave 2:  2.2006 1.4930 0.2474 0.0976

Peak and trough times and signals:
             t.Upper Z.Upper t.Lower Z.Lower
FMM wave 1:   0.4339  3.9237  5.7250  0.0146
FMM wave 2:   5.9656  0.1348  4.6103 -4.0110

Residuals:
    Min.  1st Qu.   Median     Mean  3rd Qu.     Max. 
-0.80445 -0.21776 -0.01272  0.00000  0.17982  0.66971 

R-squared:
Wave 1 Wave 2  Total 
0.6907 0.2778 0.9685 
\end{lstlisting}

The FMM wave parameter estimates, as well as the peak and trough times, together with the signal values at those times, are presented in tabular form, where each row corresponds to a component and each column to an FMM wave parameter. As part of the summary, a brief description of the residuals, the proportion of variance explained by each FMM component and by the global model are also shown.   
From the above summary results, we can see that the variance explained by the fitted model is $96.85\%$ and that the FMM waves are labelled in decreasing order according to the $R^2$ value: the explained variance is $69.07\%$ and $27.78\%$ by ``FMM wave 1'' and ``FMM wave 2'', respectively. The $\mathrm{summary}$ output can be assigned to an object to get a `list' of all the displayed results.

Other options to return the results are the functions \textit{coef()}, \textit{getFMMPeaks()} and \textit{resid()}. The first two return a `list' similar to those obtained with \textit{summary()}. The \textit{resid()} method can be used to obtain the complete residuals vector. In addition, the fitted values can be extracted by the function \textit{fitted()}, which returns a `data.frame' with two columns: time points and fitted values.
\begin{lstlisting}
R> fitted.fmm2 <- fitted(fit.fmm2)
R> head(fitted.fmm2)

  timePoints   fitted
1 0.00000000 1.251937
2 0.06346652 1.726570
3 0.12693304 2.269928
4 0.19039955 2.826444
5 0.25386607 3.319683
6 0.31733259 3.681024
\end{lstlisting}

\begin{figure}[h!]
	\centering
	\includegraphics[width=1\textwidth]{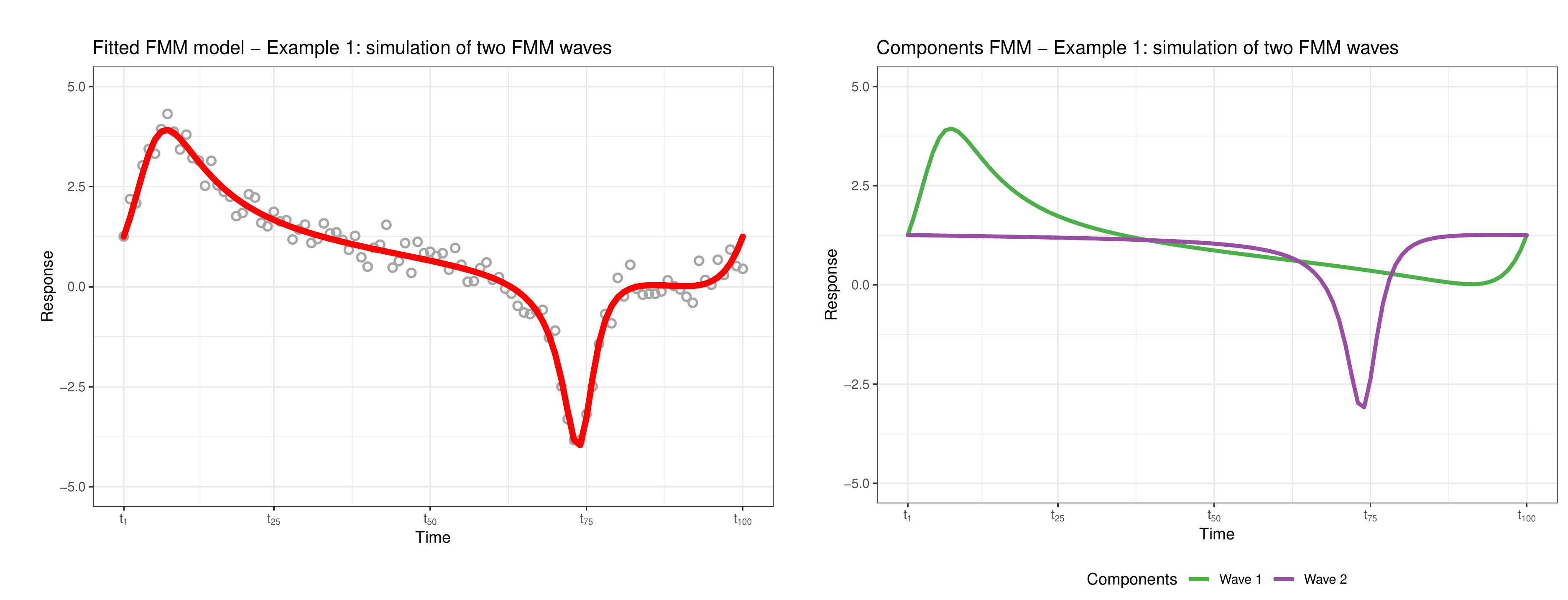}
	\caption{Graphical representations of the FMM$_2$ fitted model from the simulated data set of Example 1. The default plot is displayed on the left and the component plot is shown on the right.}
	\label{f:sim1}
\end{figure}

The code given below creates Figure~\ref{f:sim1}.

\begin{lstlisting}
R> library("RColorBrewer")
R> library("ggplot2")
R> library("gridExtra")
R> titleText <- "Example 1: two simulated FMM waves"
R> defaultFMM2 <- plotFMM(fit.fmm2, use_ggplot2 = TRUE ,textExtra = titleText) +
                  theme(plot.margin=unit(c(1,0.25,1.3,1), "cm")) + 
				  ylim(-5, 5)
R> compFMM2 <- plotFMM(fit.fmm2, components = TRUE, use_ggplot2 = TRUE ,
                       textExtra = titleText) +
               theme(plot.margin=unit(c(1,0.25,0,1), "cm")) +
               ylim(-5, 5) +
               scale_color_manual(values = brewer.pal("Set1",n = 8)[3:4])
R> grid.arrange(defaultFMM2, compFMM2, nrow = 1)
\end{lstlisting}
The FMM plots can be written in the R \textbf{graphics} or \textbf{ggplot2} packages. In this example, we use $\mathrm{use\_ggplot2 = TRUE}$ to build Figure~\ref{f:sim1} based on \textbf{ggplot2}. The use of \textbf{ggplot2} makes it easier to customize our plots and modify features, such as scales, margins, axes, etc. 
In Figure~\ref{f:sim1}, the two possible FMM plots are arranged via the $\mathrm{grid.arrange}$ function of the \textbf{gridExtra} package \citep{gridExtraPackage}. 

\subsection{Example 2: A restricted multicomponent FMM data set}\label{subsec:use:restMultFMM}
A set of $100$ observations is generated from an FMM$_4$ model with intercept parameter $M = 3$, amplitude parameters: $ A_1 = 4.5$, $A_2 = 3$, $A_3 = 1$ and $A_4 = 1.5$, and phase translation parameters: $\alpha_1 = 1.5$, $\alpha_2 = 4.2$, $\alpha_3 = 2$ and $\alpha_4 = 4.7$. The symmetry parameters are equal in all four FMM waves and, with regard to the kurtosis, pairs of waves are equal. Specifically, the shape parameters satisfy:
\begin{eqnarray*}
  &\beta_1 = \beta_2 = \beta_3 = \beta_4 = 3 \\
  &\omega_1 =\omega_2 = 0.01\\
  &\omega_3 =\omega_4 = 0.15\\
\end{eqnarray*}

Using the function \textit{generateFMM()}:
\begin{lstlisting}
R> set.seed(1115)
R> rfmm.data <- generateFMM(M = 3, A = c(4.5,3,1,1.5), alpha = c(1.5,4.2,2,4.7), 
                            beta = rep(3,4), omega = c(rep(0.01,2),rep(0.15,2)),
                            plot = FALSE, outvalues = TRUE,
                            sigmaNoise = 0.3)
\end{lstlisting}

The $\mathrm{betaRestrictions}$ and $\mathrm{omegaRestrictions}$ parameters allow a wide variety of shape restrictions to be incorporated into the fitting procedure. In this example, to impose the shape restrictiction on the fitting process, we use $\mathrm{betaRestrictions = c(1, 1, 1, 1)}$ and $\mathrm{omegaRestrictions = c(1, 1, 2, 2)}$.

\begin{figure}[!ht]
	\centering
	\includegraphics[width=1\textwidth]{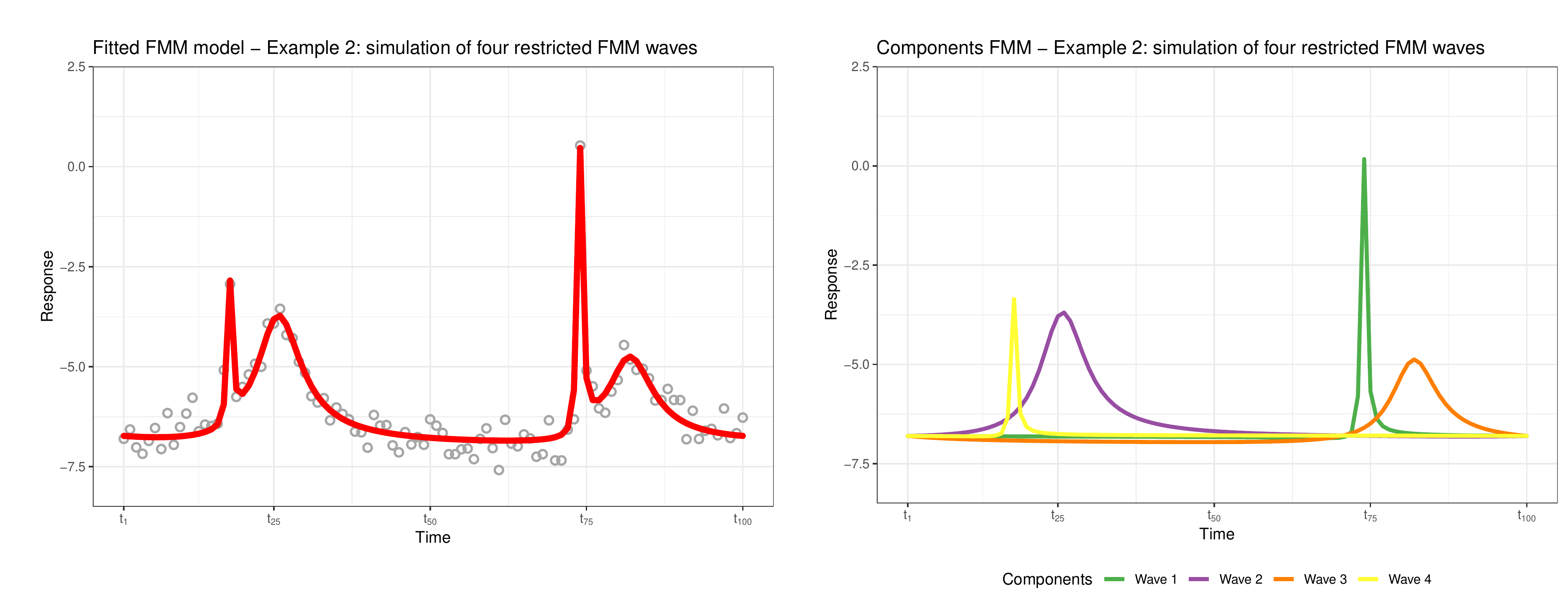}
	\caption{Graphical representation of the fitted restricted FMM$_4$ model with $\beta_1 = \beta_2 = \beta_3 = \beta_4$, $\omega_1 = \omega_3$ and $\omega_2 = \omega_4$ constraints. A scatter plot of the simulated data along with the fitted model is displayed on the left (default plot). The component plot is shown on the right.}
	\label{f:sim2}
\end{figure}

\begin{lstlisting}
R> fit.rfmm <- fitFMM(vData = rfmm.data$y, timePoints = rfmm.data$t, nback = 4,
                      betaRestrictions = c(1, 1, 1, 1),
                      omegaRestrictions = c(1, 1, 2, 2),
                      showTime = FALSE)

|--------------------------------------------------|
|==================================================|
Stopped by reaching maximum iterations ( 4 iterations ) 

R> summary(fit.rfmm)

Title:
FMM model with 4 components

Coefficients:
M (Intercept): 1.4781
                  A  alpha   beta  omega
FMM wave 1:  3.9422 1.4806 2.9996 0.0126
FMM wave 2:  1.5654 4.6953 2.9996 0.1366
FMM wave 3:  1.0413 1.9800 2.9996 0.1366
FMM wave 4:  1.8896 4.2109 2.9996 0.0126

Peak and trough times and signals:
             t.Upper Z.Upper t.Lower Z.Lower
FMM wave 1:   4.6240  1.3097  4.2721 -6.8088
FMM wave 2:   1.5732 -3.7161  5.6553 -6.3382
FMM wave 3:   5.1411 -4.7470  2.9400 -6.7395
FMM wave 4:   1.0711 -2.5415  0.7192 -6.6951

Residuals:
     Min.   1st Qu.    Median      Mean   3rd Qu.      Max. 
-0.735225 -0.217362  0.004187  0.000000  0.186612  0.928811 

R-squared:
Wave 1 Wave 2 Wave 3 Wave 4  Total 
0.3664 0.2765 0.1656 0.1081 0.9166
\end{lstlisting}
Figure \ref{f:sim2} displays the two graphical representations of the fitted model: the fitted FMM model and the component FMM plot on the left and right, respectively.

The fitting of more complex signals requires more execution time.  
By combining the values of the arguments $\mathrm{lengthAlphaGrid}$, $\mathrm{lengthOmegaGrid}$ and $\mathrm{numReps}$ of \textit{fitFMM()}, a balance between the computational cost and the accuracy of the estimates can be achieved. In this example, we managed to reduce the execution time by setting half-length sequences of the default ones, for both the $\mathrm{lengthAlphaGrid}$ and $\mathrm{lengthOmegaGrid}$ arguments, while increasing the number of repetitions from $3$ to $5$. Note that this configuration even allows us to slightly improve the accuracy of the solution from $R^2 = 91.66\%$ to $R^2 = 92.03\%$. 

\begin{lstlisting}
R> fit.rfmm.eff <- fitFMM(vData = rfmm.data$y, timePoints = rfmm.data$t, nback = 4,
                   betaRestrictions = c(1, 1, 1, 1), 
                   omegaRestrictions = c(1, 1, 2, 2),
                   showTime = FALSE,
                   lengthAlphaGrid = 24, lengthOmegaGrid = 12, numReps = 5)

|--------------------------------------------------|
|==================================================|
Stopped by reaching maximum iterations ( 4 iterations ) 

R> summary(fit.rfmm.eff)

Title:
FMM model with 4 components

Coefficients:
M (Intercept): 0.9556
                  A  alpha   beta  omega
FMM wave 1:  3.5811 1.4855 3.0122 0.0136
FMM wave 2:  1.4826 4.6988 3.0122 0.1761
FMM wave 3:  1.0190 2.0158 3.0122 0.1761
FMM wave 4:  1.9404 4.2063 3.0122 0.0136

Peak and trough times and signals:
             t.Upper Z.Upper t.Lower Z.Lower
FMM wave 1:   4.6289  0.6142  4.2143 -6.8739
FMM wave 2:   1.5800 -3.9989  5.4040 -5.5034
FMM wave 3:   5.1802 -4.9250  2.7210 -6.7180
FMM wave 4:   1.0664 -2.3001  0.6519 -6.7150

Residuals:
    Min.  1st Qu.   Median     Mean  3rd Qu.     Max. 
-0.65126 -0.20547 -0.02922  0.00000  0.20787  0.91315 

R-squared:
Wave 1 Wave 2 Wave 3 Wave 4  Total 
0.3656 0.2605 0.1822 0.1120 0.9203 
\end{lstlisting}
%

\section{Real data analysis using the \textbf{FMM} package}\label{sec:realExamp}
This section illustrates the use of the \textbf{FMM} package on the analysis of real signals from chronobiology, electrocardiography and neuroscience. To do this, the package includes four real-world data sets in $\mathrm{RData}$ format which are described in the following sections.

\subsection{Example 1: Chronobiology}\label{subsec:exCrono}
Chronobiology studies ubiquitous daily variations found in nature and in many aspects of the physiology of human beings, such as blood pressure or hormone levels \citep{Mermet+Yeung+Naef:2017}. These phenomena commonly display signals with oscillatory patterns that repeat every 24 hours, usually known as circadian rhythms. In particular, circadian gene expression data have been deeply analyzed in the literature as they regulate the vast majority of molecular rhythms involved in diverse biochemical and cellular functions, see among others \cite{Zhang+Lahens+Ballance+Hughes+Hogenesch:2014}, \cite{Cornelissen:2014} and \cite{Larriba+Rueda+Fernandez+Peddada:2020}.

The \textbf{FMM} package includes a data set called $\mathrm{mouseGeneExp}$ that contains expression data of the \textit{Iqgap2} gene from mouse liver. The liver circadian database is widely extended in chronobiology since the liver is a highly rhythmic organ with moderate levels of noise \citep{Anafi+Francey+Hogenesch+Kim:2017,Larriba+Rueda+Fernandez+Peddada:2018,Larriba+Rueda+Fernandez+Peddada:2020}. The complete database is freely available at NCBI GEO (\url{http://www.ncbi.nlm.nih.gov/geo/}), with GEO accession number GSE11923. Gene expression values are given along 48 hours with a sampling frequency of 1hour/2days. Hence, data are collected along two periods, and an FMM$_1$ model is fitted to the \textit{Iqgap2} average expressed values as follows: 
\begin{lstlisting}
R> data("mouseGeneExp", package = "FMM")
R> fitGene <- fitFMM(mouseGeneExp, nPeriods = 2, showProgress = FALSE)
R> summary(fitGene)

Title:
FMM model with 1 components

Coefficients:
M (Intercept): 10.1508
                  A  alpha   beta  omega
FMM wave 1:  0.4683 3.0839 1.5329 0.0816

Peak and trough times and signals:
             t.Upper Z.Upper t.Lower Z.Lower
FMM wave 1:   0.1115 10.6191  6.0686  9.6825

Residuals:
      Min.    1st Qu.     Median       Mean    3rd Qu.       Max. 
-9.751e-02 -3.490e-02  2.269e-03 -1.530e-06  2.670e-02  1.890e-01 

R-squared:
[1] 0.8752
\end{lstlisting}

\begin{figure}[h!]
	\centering
	\includegraphics[width=0.9\textwidth]{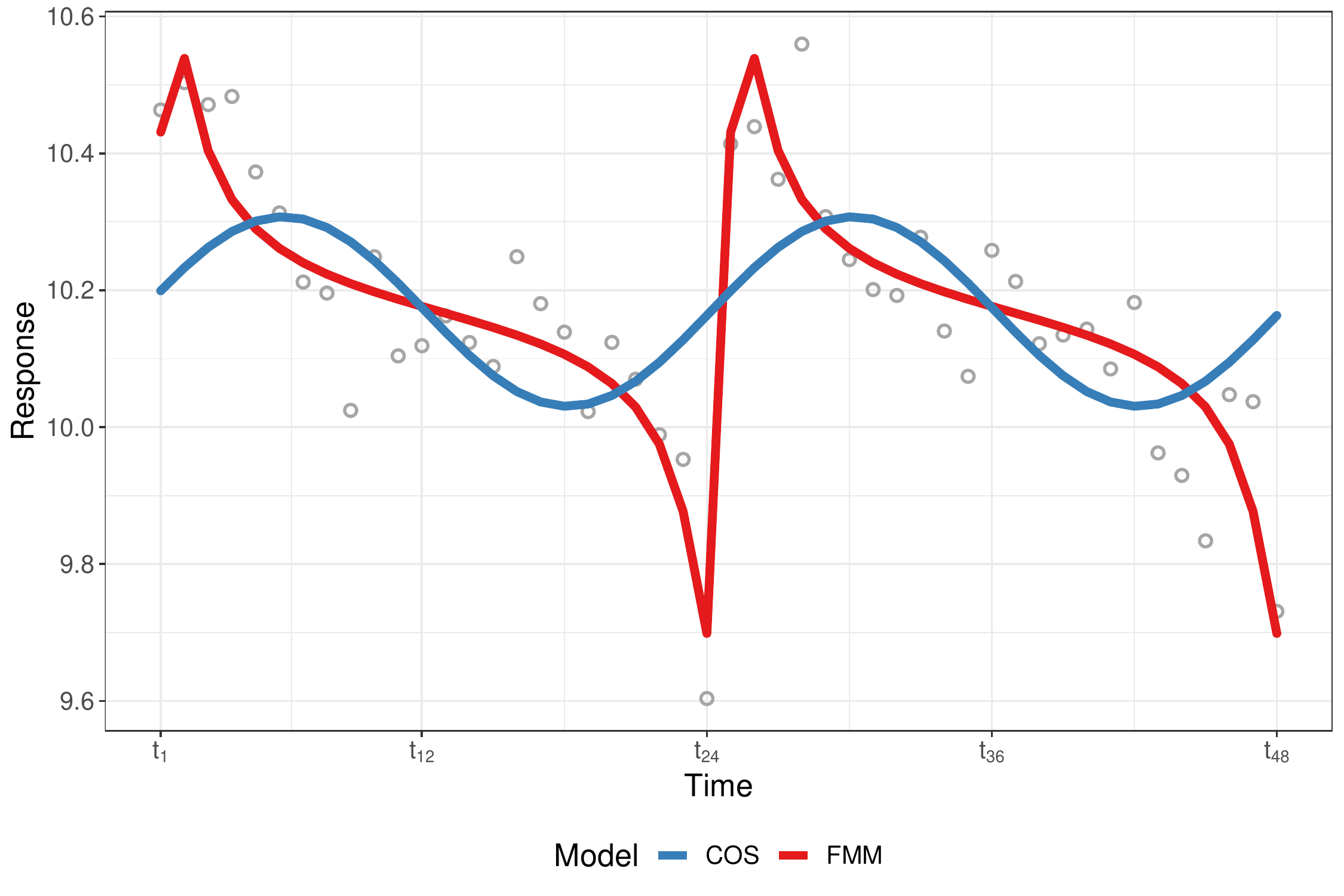}
	\caption{\textit{Iqgap2} gene expression data along two periods (grey dots); FMM (red line) and COS (blue line) fitted models.}
	\label{f:chronobiology}
\end{figure}

The behavior of the FMM versus COS model to describe this asymmetric pattern has been compared in terms of $R^2$. The FMM model clearly outperforms the COS one with an $R^2$ of $0.8752$ and $0.2835$, respectively. In addition, a difference of $4.73$ hours in peak time estimation between both models is observed, the FMM peak estimate being much more reliable, as is shown in Figure~\ref{f:chronobiology}.

\subsection{Example 2: Electrocardiography}\label{subsec:exECG}
ECG records the periodic electrical activity of the heart. This activity represents the contraction and relaxation of the atria and ventricle, processes related to the crests and troughs of the ECG waveform. Heartbeats are decomposed into five fundamental waves, labelled as $P$, $Q$, $R$, $S$ and $T$, corresponding to the different phases of the heart's electric activity. The main features used in medical practice for cardiovascular pathology diagnosis are related to the location and amplitudes of these waves, and, of them, those labeled as $P$, $R$ and $T$ are of particular interest \citep{Bayes:2007}. Standard ECG signals are registered using twelve leads, calculated from different electrode locations, Lead II being the reference one.

The \textbf{FMM} package includes the analysis of a typical ECG beat from the QT database \citep{Laguna+Mark+Goldberg+Moody+1997}. This recording, from the subject \textit{sel100}, belongs to the \textit{Normal} category, regarding Physionet's  pathology classification \citep{Goldberger+Amaral+Glass+Hausdorff+etal:2000}. The data illustrate the voltage of the heart's electric activity, measured in $mV$, along the beat with a sampling frequency of $250Hz$. Specifically, the ECG signal from Lead II in the fifth of the thirty annotated beats is analysed. Recordings are publicly available on (\url{http://www.physionet.org}). Data are saved as $\mathrm{ecgData}$ in the package. For an ECG beat, an FMM$_{ecg}$, a fifth order multicomponent FMM model can be fitted with the instruction: 
\begin{lstlisting}
R> data("ecgData", package = "FMM")
R> fitEcg <- fitFMM(ecgData, nback = 5, showProgress = FALSE)
R> summary(fitEcg)

Title:
FMM model with 5 components

Coefficients:
M (Intercept): 5.2908
                  A  alpha   beta  omega
FMM wave 1:  0.6387 5.5134 3.2786 0.0322
FMM wave 2:  0.0995 4.4191 3.7638 0.1365
FMM wave 3:  0.2412 5.3504 0.6575 0.0320
FMM wave 4:  0.3310 5.5906 4.7894 0.0122
FMM wave 5:  0.0668 1.7998 2.1309 0.1641

Peak and trough times and signals:
             t.Upper Z.Upper t.Lower Z.Lower
FMM wave 1:   2.3674  6.2328  3.2488  4.7246
FMM wave 2:   1.1898  4.9486  2.0805  4.6806
FMM wave 3:   2.3959  6.0793  2.1870  4.5543
FMM wave 4:   2.4227  5.7845  2.4715  4.7177
FMM wave 5:   5.1224  4.8646  4.3649  4.7226

Residuals:
      Min.    1st Qu.     Median       Mean    3rd Qu.       Max. 
-0.0694805 -0.0096146 -0.0001619  0.0000000  0.0098442  0.0625571 

R-squared:
Wave 1 Wave 2 Wave 3 Wave 4 Wave 5  Total 
0.7650 0.0927 0.0559 0.0402 0.0380 0.9919 
\end{lstlisting}

\begin{figure}[htbp]
\includegraphics[width=1\textwidth]{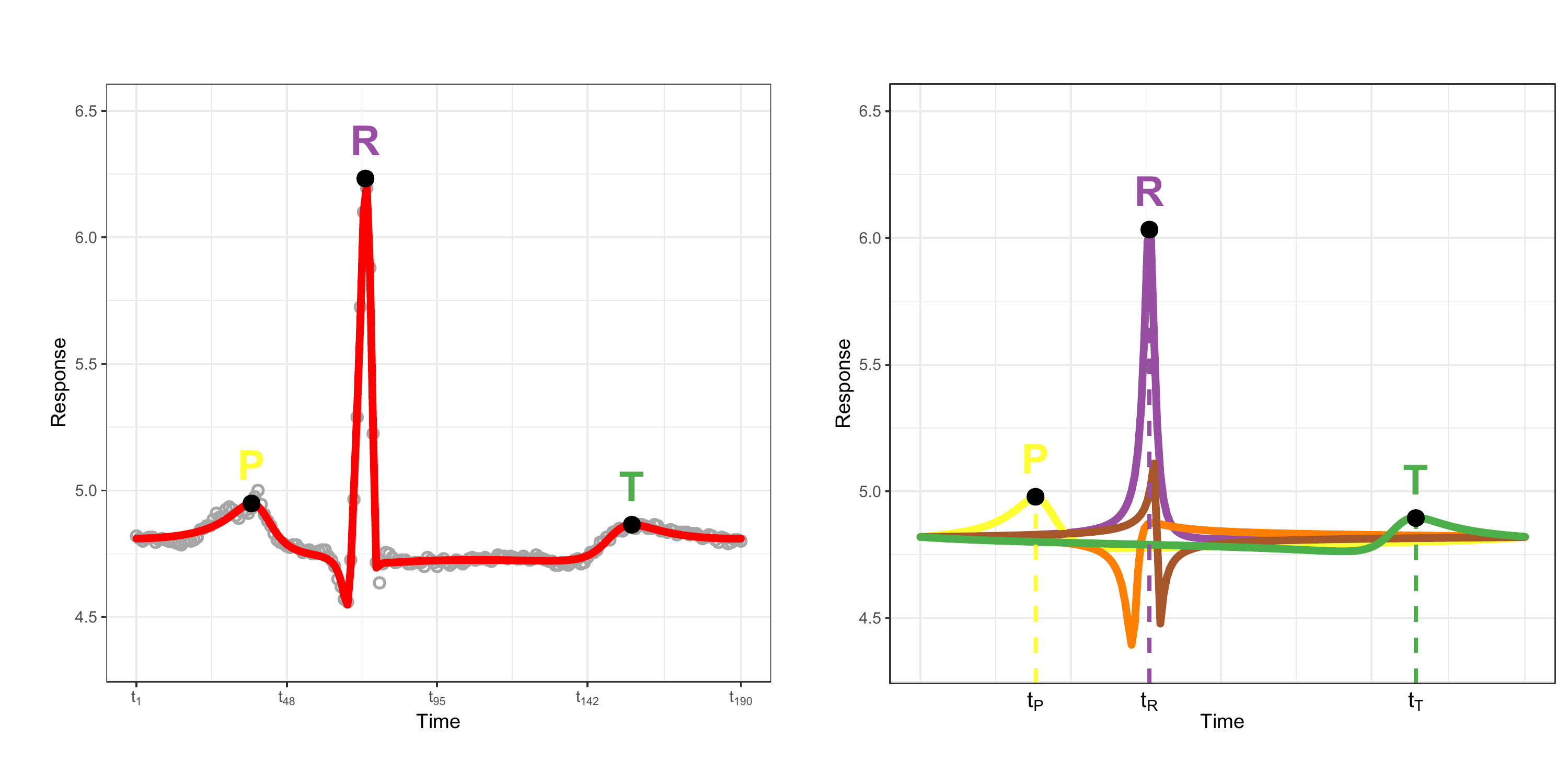}
      \caption{FMM$_{ecg}$ performance on a single beat from patient \textit{sel100} from the QT database. Left: Data (grey dots) and FMM fitting (red line). Black dots locate the $P$, $R$ and $T$ fiducial marks. Right: ECG decomposition on $P$(orange), $Q$ (purple), $R$ (green), $S$ (yellow) and $T$ (blue) waves. Dash lines indicate $P$, $R$ and $T$ peak times.} \label{f:ecg}
\end{figure}

It is worth noting that the \textbf{FMM} package not only provides ECG signal-fitting (the left hand panel in Figure~\ref{f:ecg}), but it also does wave decomposition and fiducial mark annotations on the desired waves (the right hand panel in Figure~\ref{f:ecg}). It is clearly visible how the specific shapes of the five main waves contribute to drawing and explaining the Lead II ECG waveform from the \textit{Normal} morphology. See \cite{Rueda+Larriba+Lamela:2021} for a complete review of FMM$_{ecg}$.

\subsection{Example 3: Neuroscience}\label{subsec:exNeuro}
\subsubsection{Single AP curve}\label{subsubsec:singleAP}
The study of the electrophysiological activity of neurons is one of the main research branches in neuroscience. The AP curves are oscillatory signals that serve as basic information units between neurons. They measure the electrical potential difference between inside and outside the cell due to an external stimulus. \cite{Gerstner+Kistler+Naud+Paninski:2014} can serve as a basic reference for electrophysiological neuroscience. Recently, the shape and other features of the AP have been used in problems such as spike sorting \citep{Racz+Liber+Nemeth+Fiath+Rokai+Harmati+Ulbert+Marton:2020, Souza+Lopes+Bacelo+Tort:2019, Caro+Delgado+Gruart+Sanchez:2018} or neuronal cell type classification \citep{Teeter+Iyer+Menon+al:2018, Gouwens+Sorense+Berg+al:2019, Mosher+Wei+Kaminski+Nandi+Mamelak+Anastassiou+Rutishauser:2020, RodriguezCollado+Rueda:2021b}.

The package includes an example of a neuronal AP. The data were simulated with the renowned Hodgkin-Huxley model, first presented in \cite{Hodgkin+Huxley:1952}, which is defined as a system of ordinary differential equations and has been used in a wide array of applications, as it successfully describes the neuronal activity in various organisms. The simulation has been done using a modified version of the python package \textbf{NeuroDynex} available at \cite{Gerstner+Kistler+Naud+Paninski:2014}. More concretely, a short square stimulus of $12 \mu A$ has been applied to the neuron.
The data can be accurately fitted by an FMM$_2$ model as follows:
\begin{lstlisting}
R> data("neuronalSpike", package = "FMM")
R> fitSingleAP <- fitFMM(neuronalSpike, nback = 2, showProgress = FALSE)
R> summary(fitSingleAP)

Title:
FMM model with 2 components

Coefficients:
M (Intercept): 44.9448
                   A  alpha   beta  omega
FMM wave 1:  52.9023 4.4160 3.0606 0.0413
FMM wave 2:  18.5044 4.6564 4.9623 0.0322

Peak and trough times and signals:
             t.Upper  Z.Upper t.Lower  Z.Lower
FMM wave 1:   1.2777 110.8354  5.9670  -2.5003
FMM wave 2:   1.4320  36.9015  1.5649 -16.2606

Residuals:
    Min.  1st Qu.   Median     Mean  3rd Qu.     Max. 
-14.3012  -1.0033   0.7473   0.0000   1.3230  24.8616 

R-squared:
Wave 1 Wave 2  Total 
0.9064 0.0604 0.9669 
\end{lstlisting}

\begin{figure}[h]
	\centering
	\includegraphics[width=0.8\textwidth]{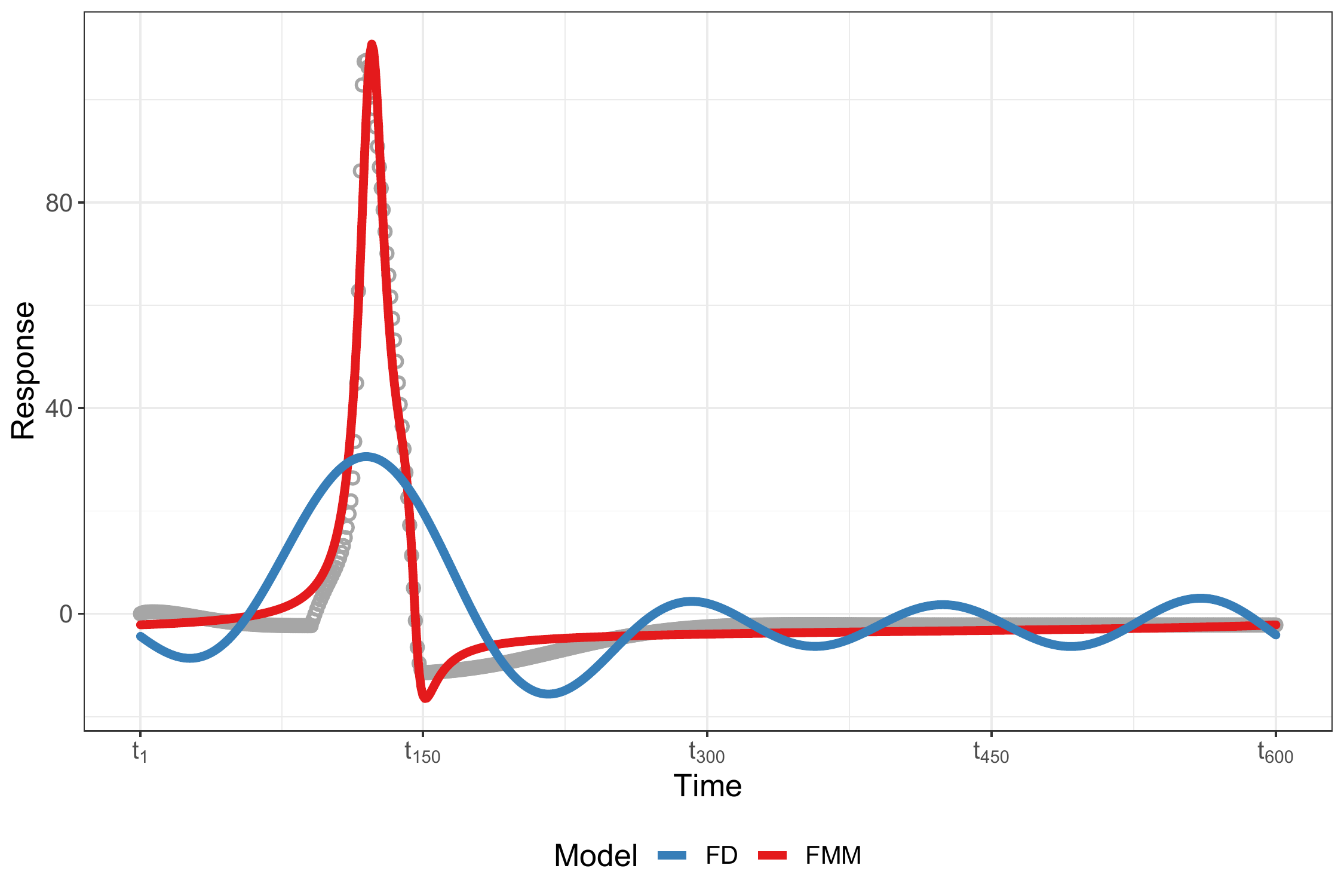}
	\caption{Neuronal AP simulated with the Hodgkin-Huxley model (parameters: $C=1, g_{Na}=260, g_{K}=30, g_{L}=0.31, V_{K}=-12, V_{Na}=115, V_{L}=10.6, \widetilde{a}_n=1.15, \widetilde{b}_n=0.85, \widetilde{a}_m=0.9, \widetilde{b}_m=1.3, \widetilde{a}_h=1, \widetilde{b}_h=1$ and applying a current of $12 \mu A$ for $1$ millisecond) and the estimated FMM$_2$ signal in red. An FD model of the same number of degree of freedom has been fitted and plotted in blue.}
	\label{f:FigureNeuro}
\end{figure}

The goodness of fit of the FMM$_2$ model can be ascertained in Figure~\ref{f:FigureNeuro}. For comparison purposes, an FD model has been fitted with the same number of degrees of freedom. While the FD attains an $R^2 = 0.3926$, the FMM model achieves a better fit with $R^2 = 0.9669$.

\subsubsection{AP train}\label{subsubsec:APtrain}
Multiple AP curves, denominated spike or AP train, are usually observed as the response to a stimulus. Various models, such as the widely used leaky-and-fire models \citep{Lynch+Houghton:2015}, cut the signal into segments, each one containing an AP curve. Some authors suggest cutting the signal into even segments \citep{Gerstner+Kistler+Naud+Paninski:2014}. However, the length of the segments turns out to be significantly different between different types of neurons, as explained in \cite{Teeter+Iyer+Menon+al:2018}, and unequal data segments can lessen the utility of some approaches. An important aspect to take into account is that the shape of the APs in the spike train is considered to be similar and, consequently, a restricted FMM model can accurately fit the entire signal.

The \textbf{FMM} package includes the data of a spike train composed of three AP curves. The proposed model for use with these data is an FMM$_{ST}$ model, as defined in \cite{RodriguezCollado+Rueda:2021}. Each AP is modeled by two components. The $\beta$ and $\omega$ parameters are constrained between AP curves. The code below fits the model. 
\begin{lstlisting}
R> data("neuronalAPTrain", package = "FMM")
R> nAPs <- 3; restriction <- c(rep(1,nAPs),rep(2,nAPs))
R> fitAPTrain<-fitFMM(neuronalAPTrain, nback = nAPs*2, 
                      betaRestrictions = restriction, 
                      omegaRestrictions = restriction, 
                      showProgress = FALSE)
R> summary(fitAPTrain)

Title:
FMM model with 6 components

Coefficients:
M (Intercept): 135.4915
                   A  alpha   beta  omega
FMM wave 1:  51.1192 4.2327 2.8190 0.0384
FMM wave 2:  51.7219 6.1358 2.8190 0.0384
FMM wave 3:  52.1105 1.7536 2.8190 0.0384
FMM wave 4:  19.6488 0.0975 4.8644 0.0551
FMM wave 5:  19.2020 1.9978 4.8644 0.0551
FMM wave 6:  20.3642 4.4783 4.8644 0.0551

Peak and trough times and signals:
             t.Upper  Z.Upper t.Lower  Z.Lower
FMM wave 1:   1.1036 111.2685  0.6274  -0.1787
FMM wave 2:   3.0067 111.4205  2.5304  -1.2404
FMM wave 3:   4.9077 111.7226  4.4314  -1.4963
FMM wave 4:   3.1109  58.2307  3.3336 -13.7196
FMM wave 5:   5.0112  58.2938  5.2339 -13.2943
FMM wave 6:   1.2086  58.5306  1.4313 -14.2673

Residuals:
    Min.  1st Qu.   Median     Mean  3rd Qu.     Max. 
-15.1339  -1.5071   0.4973   0.0000   1.6303  19.9627 

R-squared:
Wave 1 Wave 2 Wave 3 Wave 4 Wave 5 Wave 6  Total 
0.3502 0.2922 0.2485 0.0411 0.0274 0.0244 0.9838
\end{lstlisting}

\begin{figure}[!ht]
\centering
\includegraphics[width=1\textwidth]{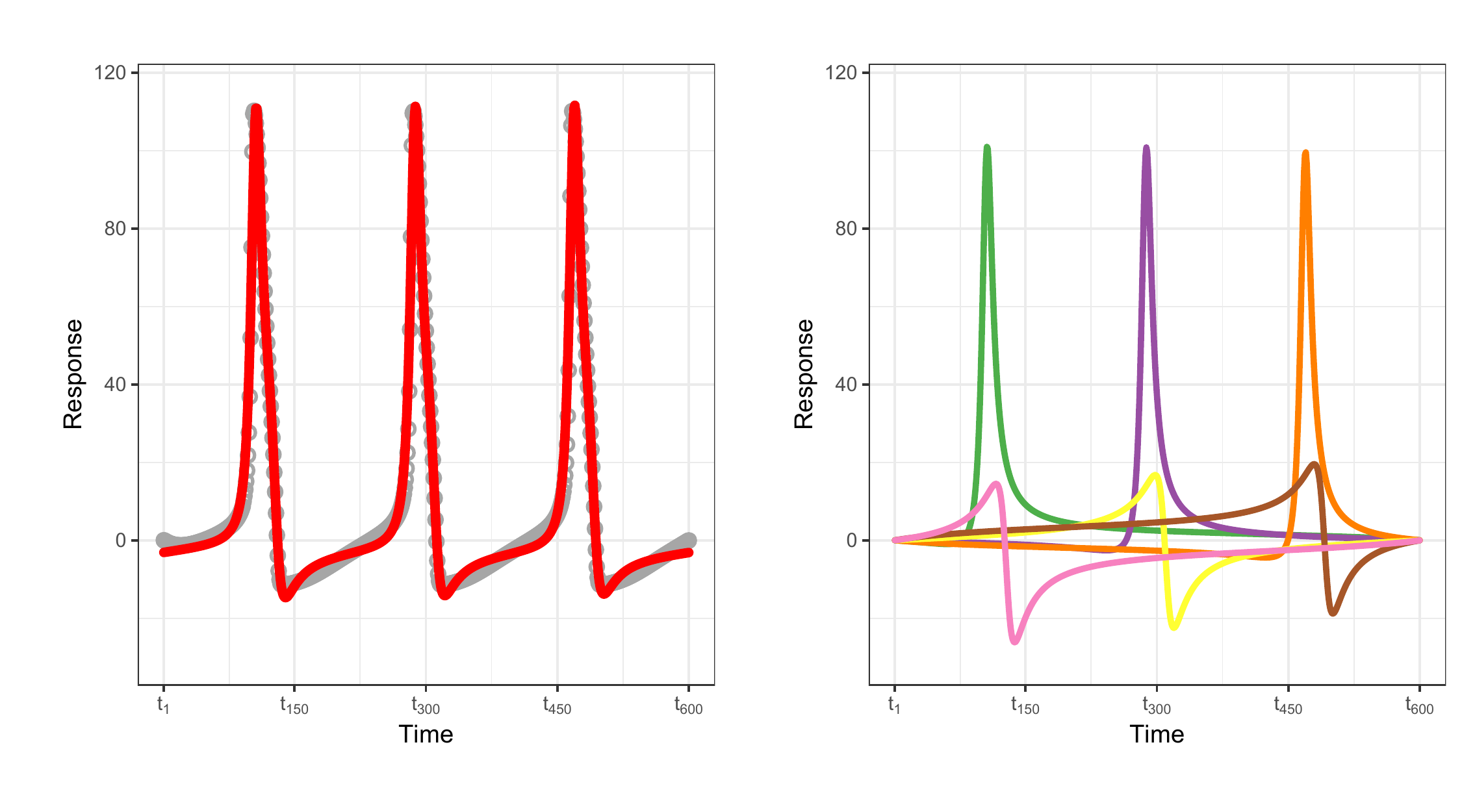}
\caption{Neuronal AP simulated with the Hodgkin-Huxley model (parameters: $C=1, g_{Na}=232, g_{K}=45, g_{L}=0.215, V_{K}=-12, V_{Na}=115, V_{L}=10.6, \widetilde{a}_n=0.95, \widetilde{b}_n=1.3, \widetilde{a}_m=1, \widetilde{b}_m=1.15, \widetilde{a}_h=1, \widetilde{b}_h=1$ and applying a short square current of 4.5 $\mu A$ for $1$ millisecond) and the estimated FMM$_{ST}$ signal in red. The components plot of the model can be seen on the right hand side of the figure.}
\label{f:FigureNeuroTrain}
\end{figure}

In Figure~\ref{f:FigureNeuroTrain}, the fit of the FMM$_{ST}$ model can be visualized. The goodness of fit of the model is excellent, achieving an $R^2 = 0.9838$.

\section{Summary} \label{sec:summary}
A general overview on the R package \textbf{FMM}, which implements the estimation of FMM models, is provided in this paper. The flexibility offered by these models to fit oscillatory signals of many different shapes makes them a very useful tool to model complex rhythmic patterns. The FMM methodology and its application to very diverse biological data has been described in previous papers \citep{Rueda+Larriba+Peddada:2019, Rueda+Larriba+Lamela:2021, Rueda+Rodriguez+Larriba:2021} and recently revised in \cite{Rueda+Fernandez+Larriba+Rodriguez:2021}.

The package allows both single and multicomponent FMM models to be estimated. In order to provide greater flexibility, equality constraints for shape parameters have also been implemented. In addition, graphical representations of the fitted models and the possibility of generating synthetic data are available. The functionality of the package has been illustrated by simulated data and also by real examples from different areas of application related to present-day biological problems.

Possible future extensions of the \textbf{FMM} package include the implementation of additional restrictions to suit the model to other real signals; the possibility to include weights that determine how much each observation influences the parameter estimates; and the improvement of the computational efficiency. Currently, one of the most important limitations of the stable version of the package (version $0.1.2$) is that the fitting process, especially of the restricted model, can be a time-consuming process. Nevertheless, the execution times of the development version available on the GitHub repository (\url{https://github.com/alexARC26/FMM}) have already been significantly reduced, in many cases by an order of $10$ times or more. 

\section*{Acknowledgments}
This research was partially supported by the Spanish Ministerio de Econom\'ia y Competitividad, grant PID2019-106363RB-I00.

\bibliographystyle{unsrtnat}


\begin{thebibliography}{45}
\newcommand{\enquote}[1]{``#1''}
\providecommand{\natexlab}[1]{#1}
\providecommand{\url}[1]{\texttt{#1}}
\providecommand{\urlprefix}{URL }
\expandafter\ifx\csname urlstyle\endcsname\relax
  \providecommand{\doi}[1]{doi:\discretionary{}{}{}#1}\else
  \providecommand{\doi}{doi:\discretionary{}{}{}\begingroup
  \urlstyle{rm}\Url}\fi
\providecommand{\eprint}[2][]{\url{#2}}

\bibitem[{Anafi \emph{et~al.}(2017)Anafi, Francey, Hogenesch, and
  Kim}]{Anafi+Francey+Hogenesch+Kim:2017}
Anafi RC, Francey LJ, Hogenesch JB, Kim J (2017).
\newblock \enquote{CYCLOPS Reveals Human Transcriptional Rhythms in Health and
  Disease.}
\newblock \emph{Proceedings of the National Academy of Sciences},
  \textbf{114}(20), 5312--5317.
\newblock \doi{10.1073/pnas.1619320114}.

\bibitem[{Auguie(2017)}]{gridExtraPackage}
Auguie B (2017).
\newblock \emph{\textbf{gridExtra}: Miscellaneous Functions for Grid Graphics}.
\newblock R package version 2.3,
  \urlprefix\url{https://CRAN.R-project.org/package=gridExtra}.

\bibitem[{{Bayes de Luna}(2007)}]{Bayes:2007}
{Bayes de Luna} A (2007).
\newblock \emph{Basic Electrocardiography: Normal and Abnormal ECG Patterns}.
\newblock John Wiley \& Sons, Ltd.
\newblock \doi{10.1002/9780470692622}.

\bibitem[{Boashash(2016)}]{Boashash:2016}
Boashash B (2016).
\newblock \emph{Time-Frequency Signal Analysis and Processing: A Comprehensive
  Reference}.
\newblock 2nd edition. Academic Press, San Francisco, CA.

\bibitem[{Carlucci \emph{et~al.}(2020)Carlucci, Kri\v{s}\v{c}i\={u}nas, Li,
  Gibas, Koncevi\v{c}ius, Petronis, and Oh}]{Calucci+Kris+Li+al:2020}
Carlucci M, Kri\v{s}\v{c}i\={u}nas A, Li H, Gibas P, Koncevi\v{c}ius K,
  Petronis A, Oh G (2020).
\newblock \enquote{\textbf{DiscoRhythm}: An Easy-To-Use Web Application And
  R Package for Discovering Rhythmicity.}
\newblock \emph{Bioinformatics}, \textbf{36}(6), 1952–--1954.
\newblock \doi{10.1093/bioinformatics/btz834}.

\bibitem[{Caro-Mart{\'\i}n \emph{et~al.}(2018)Caro-Mart{\'\i}n,
  Delgado-Garc{\'\i}a, Gruart, and
  S{\'a}nchez-Campusano}]{Caro+Delgado+Gruart+Sanchez:2018}
Caro-Mart{\'\i}n, Delgado-Garc{\'\i}a, Gruart, S{\'a}nchez-Campusano (2018).
\newblock \enquote{Spike Sorting Based on Shape, Phase, and Distribution
  Features, and K-TOPS Clustering with Validity and Error Indices.}
\newblock \emph{Scientific Reports}, \textbf{8}(1), 1--28.
\newblock \doi{10.1038/s41598-018-35491-4}.

\bibitem[{Chang \emph{et~al.}(2020)Chang, Cheng, Allaire, Xie, and
  McPherson}]{Chang+Cheng+Allaire+Xie+McPherson:2020}
Chang W, Cheng J, Allaire J, Xie Y, McPherson J (2020).
\newblock \emph{\textbf{shiny}: Web Application Framework for R}.
\newblock R package version 1.5.0,
  \urlprefix\url{https://CRAN.R-project.org/package=shiny}.

\bibitem[{Cornelissen(2014)}]{Cornelissen:2014}
Cornelissen G (2014).
\newblock \enquote{Cosinor-Based Rhythmometry.}
\newblock \emph{Theoretical Biology and Medical Modelling}, \textbf{11}, 16.
\newblock \doi{10.1186/1742-4682-11-16}.

\bibitem[{Corporation and Weston(2020)}]{Weston:2020}
Corporation M, Weston S (2020).
\newblock \emph{\textbf{doParallel}: Foreach Parallel Adaptor for the 'parallel'
  Package}.
\newblock R package version 1.0.16,
  \urlprefix\url{https://CRAN.R-project.org/package=doParallel}.

\bibitem[{Downs and Mardia(2002)}]{Downs+Mardia:2002}
Downs TD, Mardia KV (2002).
\newblock \enquote{Circular Regression.}
\newblock \emph{Biometrika}, \textbf{89}(3), 683--698.
\newblock \doi{10.1093/biomet/89.3.683}.

\bibitem[{Gerstner \emph{et~al.}(2014)Gerstner, Kistler, Naud, and
  Paninski}]{Gerstner+Kistler+Naud+Paninski:2014}
Gerstner W, Kistler WM, Naud R, Paninski L (2014).
\newblock \emph{Neuronal Dynamics: From Single Neurons to Networks and Models
  of Cognition}.
\newblock Cambridge University Press.
\newblock \doi{10.1017/CBO9781107447615}.

\bibitem[{Gierke \emph{et~al.}(2018)Gierke, Helget, and
  Cornelissen-Guillaume}]{Gierke+Helget+Cronelissen:2018}
Gierke CL, Helget R, Cornelissen-Guillaume G (2018).
\newblock \emph{\textbf{CATkit}: Chronomics Analysis Toolkit (CAT): Periodicity
  Analysis}.
\newblock R package version 3.3.3,
  \urlprefix\url{https://CRAN.R-project.org/package=CATkit}.

\bibitem[{Goldberger \emph{et~al.}(2000)Goldberger, Amaral, Glass, Hausdorff,
  Ivanov, Mark, Mietus, Moody, Peng, and
  Stanley}]{Goldberger+Amaral+Glass+Hausdorff+etal:2000}
Goldberger AL, Amaral LA, Glass L, Hausdorff JM, Ivanov PC, Mark RG, Mietus JE,
  Moody GB, Peng CK, Stanley HE (2000).
\newblock \enquote{PhysioBank, PhysioToolkit, and PhysioNet: Components of a
  New Research Resource for Complex Physiologic Signals.}
\newblock \emph{Circulation}, \textbf{101}(23), E215--20.
\newblock \doi{10.1161/01.cir.101.23.e215}.

\bibitem[{Gouwens \emph{et~al.}(2019)Gouwens, Sorense, Berg, Lee
  \emph{et~al.}}]{Gouwens+Sorense+Berg+al:2019}
Gouwens NW, Sorense SA, Berg J, Lee C, \emph{et~al.} (2019).
\newblock \enquote{Classification of Electrophysiological and Morphological
  Neuron Types in the Mouse Visual Cortex.}
\newblock \emph{Nature Neuroscience}, \textbf{22}, 1182–--1195.
\newblock \doi{10.1038/s41593-019-0417-0}.

\bibitem[{Hodgkin and Huxley(1952)}]{Hodgkin+Huxley:1952}
Hodgkin AL, Huxley AF (1952).
\newblock \enquote{A Quantitative Description of Membrane Current and Its
  Application to Conduction and Excitation in Nerve.}
\newblock \emph{The Journal of Physiology}, \textbf{117}(4), 500--544.
\newblock \doi{10.1113/jphysiol.1952.sp004764}.

\bibitem[{Hughes \emph{et~al.}(2010)Hughes, Hogenesch, and
  Kornacker}]{Hughes+Hogenesch+Kornacker:2010}
Hughes ME, Hogenesch JB, Kornacker K (2010).
\newblock \enquote{JTK\_CYCLE: An Efficient Nonparametric Algorithm for
  Detecting Rhythmic Components in Genome Scale Data Sets.}
\newblock \emph{Journal of Biological Rhythms}, \textbf{25}(5), 372--380.
\newblock \doi{10.1177/0748730410379711}.

\bibitem[{Kato \emph{et~al.}(2008)Kato, Shimizu, and
  Shieh}]{Kato+Shimizu+Shieh:2008}
Kato S, Shimizu K, Shieh GS (2008).
\newblock \enquote{A Circular - Circular Regression Model.}
\newblock \emph{Statistica Sinica}, \textbf{18}(2), 633--645.

\bibitem[{Laguna \emph{et~al.}(1997)Laguna, Mark, Goldberg, and
  Moody}]{Laguna+Mark+Goldberg+Moody+1997}
Laguna P, Mark RG, Goldberg A, Moody GB (1997).
\newblock \enquote{A Database for Evaluation of Algorithms for Measurement of
  QT and Other Waveform Intervals in the ECG.}
\newblock In \emph{Computers in Cardiology 1997}, pp. 673--676. IEEE.

\bibitem[{Larriba \emph{et~al.}(2018)Larriba, Rueda, Fern\'andez, and
  Peddada}]{Larriba+Rueda+Fernandez+Peddada:2018}
Larriba Y, Rueda C, Fern\'andez MA, Peddada SD (2018).
\newblock \enquote{A Bootstrap Based Measure Robust to the Choice of
  Normalization Methods for Detecting Rhythmic Features in High Dimensional
  Data.}
\newblock \emph{Frontiers in Genetics}, \textbf{9}, 24.
\newblock \doi{10.3389/fgene.2018.00024}.

\bibitem[{Larriba \emph{et~al.}(2020)Larriba, Rueda, Fern\'andez, and
  Peddada}]{Larriba+Rueda+Fernandez+Peddada:2020}
Larriba Y, Rueda C, Fern\'andez MA, Peddada SD (2020).
\newblock \enquote{Order Restricted Inference in Chronobiology.}
\newblock \emph{Statistics in Medicine}, \textbf{39}(3), 265--278.
\newblock \doi{10.1002/sim.8397}.

\bibitem[{Lynch and Houghton(2015)}]{Lynch+Houghton:2015}
Lynch EP, Houghton CJ (2015).
\newblock \enquote{Parameter Estimation of Neuron Models Using In-Vitro and
  In-Vivo Electrophysiological Data.}
\newblock \emph{Frontiers in Neuroinformatics}, \textbf{9}, 10.
\newblock \doi{10.3389/fninf.2015.00010}.

\bibitem[{Mermet \emph{et~al.}(2017)Mermet, Yeung, and
  Naef}]{Mermet+Yeung+Naef:2017}
Mermet J, Yeung J, Naef F (2017).
\newblock \enquote{Systems Chronobiology: Global Analysis of Gene Regulation in
  a 24-Hour Periodic World.}
\newblock \emph{Cold Spring Harbor Perspectives in Biology}, \textbf{9}(3).
\newblock \doi{10.1101/cshperspect.a028720}.

\bibitem[{Mosher \emph{et~al.}(2020)Mosher, Wei, Kami\'nski, Nandi, Mamelak,
  Anastassiou, and
  Rutishauser}]{Mosher+Wei+Kaminski+Nandi+Mamelak+Anastassiou+Rutishauser:2020}
Mosher CP, Wei Y, Kami\'nski J, Nandi A, Mamelak AN, Anastassiou CA,
  Rutishauser U (2020).
\newblock \enquote{Cellular Classes in the Human Brain Revealed In Vivo by
  Heartbeat-Related Modulation of the Extracellular Action Potential Waveform.}
\newblock \emph{Cell Reports}, \textbf{30}(10), 3536--3551.e6.
\newblock \doi{10.1016/j.celrep.2020.02.027}.

\bibitem[{Moskon(2020)}]{Moskon:2020}
Moskon M (2020).
\newblock \enquote{\textbf{CosinorPy}: A python Package for
  Cosinor-Based Rhythmometry.}
\newblock \emph{BMC Bioinformatics}, \textbf{21}(1), 485.
\newblock \doi{10.1186/s12859-020-03830-w}.

\bibitem[{Mutak(2018)}]{Mutak:2018}
Mutak A (2018).
\newblock \emph{\textbf{cosinor2}: Extended Tools for Cosinor Analysis of
  Rhythms}.
\newblock R package version 0.2.1,
  \urlprefix\url{https://CRAN.R-project.org/package=cosinor2}.

\bibitem[{Nelder and Mead(1965)}]{Nelder+Mead:1965}
Nelder JA, Mead R (1965).
\newblock \enquote{A Simplex Method for Function Minimization.}
\newblock \emph{The Computer Journal}, \textbf{7}(4), 308--313.
\newblock \doi{10.1093/comjnl/7.4.308}.

\bibitem[{Neuwirth(2014)}]{Neuwirth:2014}
Neuwirth E (2014).
\newblock \emph{\textbf{RColorBrewer}: ColorBrewer Palettes}.
\newblock R package version 1.1-2,
  \urlprefix\url{https://CRAN.R-project.org/package=RColorBrewer}.

\bibitem[{Parsons \emph{et~al.}(2019)Parsons, Parsons, Garner, Oster, and
  Rawashdeh}]{Parsons+Garner+Oster+Rawashdeh:2019}
Parsons R, Parsons R, Garner N, Oster H, Rawashdeh O (2019).
\newblock \enquote{\textbf{CircaCompare}: A Method to Estimate and Statistically
  Support Differences in Mesor, Amplitude and Phase, Between Circadian
  Rhythms.}
\newblock \emph{Bioinformatics}, \textbf{36}(4), 1208--1212.
\newblock \doi{10.1093/bioinformatics/btz730}.

\bibitem[{R{\'a}cz \emph{et~al.}(2020)R{\'a}cz, Liber, N{\'e}meth, Fi{\'a}th,
  Rokai, Harmati, Ulbert, and
  M{\'a}rton}]{Racz+Liber+Nemeth+Fiath+Rokai+Harmati+Ulbert+Marton:2020}
R{\'a}cz M, Liber C, N{\'e}meth E, Fi{\'a}th R, Rokai J, Harmati I, Ulbert I,
  M{\'a}rton G (2020).
\newblock \enquote{Spike Detection and Sorting with Deep Learning.}
\newblock \emph{Journal of Neural Engineering}, \textbf{17}(1), 016038.
\newblock \doi{10.1088/1741-2552/ab4896}.

\bibitem[{{R Core Team}(2020)}]{R}
{R Core Team} (2020).
\newblock \emph{R: A Language and Environment for Statistical
  Computing}.
\newblock R Foundation for Statistical Computing, Vienna, Austria.
\newblock \urlprefix\url{https://www.R-project.org/}.

\bibitem[{Revelle(2018)}]{Revelle:2018}
Revelle W (2018).
\newblock \emph{\textbf{psych}: Procedures for Psychological, Psychometric, and
  Personality Research}.
\newblock R package version 1.8.12,
  \urlprefix\url{https://CRAN.R-project.org/package=psych}.

\bibitem[{Rodr{\'\i}guez-Collado and
  Rueda(2021{\natexlab{a}})}]{RodriguezCollado+Rueda:2021b}
Rodr{\'\i}guez-Collado A, Rueda C (2021{\natexlab{a}}).
\newblock \enquote{Electrophysiological and Transcriptomic Features Reveal a
  Circular Taxonomy of Cortical Neurons.}
\newblock \doi{10.1101/2021.03.24.436849}.
\newblock Preprint on webpage at
  \url{https://www.biorxiv.org/content/10.1101/2021.03.24.436849v3}.

\bibitem[{Rodr{\'\i}guez-Collado and
  Rueda(2021{\natexlab{b}})}]{RodriguezCollado+Rueda:2021}
Rodr{\'\i}guez-Collado A, Rueda C (2021{\natexlab{b}}).
\newblock \enquote{A Simple Parametric Representation of the Hodgkin-Huxley
  Model.}
\newblock \doi{10.1101/2021.01.11.426189}.
\newblock Preprint on webpage at
  \url{https://www.biorxiv.org/content/10.1101/2021.01.11.426189v1}.

\bibitem[{Rueda \emph{et~al.}(2021{\natexlab{a}})Rueda, Fernandez, Larriba, and
  Rodriguez-Collado}]{Rueda+Fernandez+Larriba+Rodriguez:2021}
Rueda C, Fernandez I, Larriba Y, Rodriguez-Collado A (2021{\natexlab{a}}).
\newblock \enquote{The FMM Approach to Analyze Biomedical Signals: Theory,
  Software, Applications and Future.}
\newblock Preprint.

\bibitem[{Rueda \emph{et~al.}(2021{\natexlab{b}})Rueda, Larriba, and
  Lamela}]{Rueda+Larriba+Lamela:2021}
Rueda C, Larriba Y, Lamela A (2021{\natexlab{b}}).
\newblock \enquote{The Hidden Waves in the ECG Uncovered Revealing a Sound
  Automated Interpretation Method.}
\newblock \emph{Scientific Reports}, \textbf{11}, 3724.
\newblock \doi{10.1038/s41598-021-82520-w}.

\bibitem[{Rueda \emph{et~al.}(2019)Rueda, Larriba, and
  Peddada}]{Rueda+Larriba+Peddada:2019}
Rueda C, Larriba Y, Peddada SD (2019).
\newblock \enquote{Frequency Modulated M\"obius Model Accurately Predicts
  Rhythmic Signals in Biological and Physical Sciences.}
\newblock \emph{Scientific Reports}, \textbf{9}(1), 18701.
\newblock \doi{10.1038/s41598-019-54569-1}.

\bibitem[{Rueda \emph{et~al.}(2021{\natexlab{c}})Rueda, Rodr\'iguez-Collado,
  and Larriba}]{Rueda+Rodriguez+Larriba:2021}
Rueda C, Rodr\'iguez-Collado A, Larriba Y (2021{\natexlab{c}}).
\newblock \enquote{A Novel Wave Decomposition for Oscillatory Signals.}
\newblock \emph{IEEE Transactionns on Signal Processing}, \textbf{69},
  960–--972.
\newblock \doi{10.1109/TSP.2021.3051428}.

\bibitem[{Sachs(2014)}]{Sachs:2014}
Sachs M (2014).
\newblock \emph{\textbf{cosinor}: Tools for Estimating and Predicting the Cosinor
  Model}.
\newblock R package version 1.1,
  \urlprefix\url{https://CRAN.R-project.org/package=cosinor}.

\bibitem[{Shah(2020)}]{cardPakage}
Shah AS (2020).
\newblock \emph{\textbf{card}: Cardiovascular and Autonomic Research Design}.
\newblock R package version 0.1.0,
  \urlprefix\url{https://CRAN.R-project.org/package=card}.

\bibitem[{Singer and Hughey(2019)}]{Singer+Hughey:2019}
Singer JM, Hughey JJ (2019).
\newblock \enquote{\textbf{LimoRhyde}: A Flexible Approach for Differential
  Analysis of Rhythmic Transcriptome Data.}
\newblock \emph{Journal of Biological Rhythms}, \textbf{34}(1), 5--18.
\newblock \doi{10.1177/0748730418813785}.

\bibitem[{Souza \emph{et~al.}(2019)Souza, dos Santos, Bacelo, and
  Tort}]{Souza+Lopes+Bacelo+Tort:2019}
Souza BC, dos Santos VL, Bacelo J, Tort AB (2019).
\newblock \enquote{Spike Sorting with Gaussian Mixture Models.}
\newblock \emph{Scientific Reports}, \textbf{9}(1), 1--14.
\newblock \doi{10.1038/s41598-019-39986-6}.

\bibitem[{Teeter \emph{et~al.}(2018)Teeter, Iyer, Menon, Gouwens, Feng, Berg,
  Szafer, Cain, Zeng, Hawrylycz, Koch, and Mihalas}]{Teeter+Iyer+Menon+al:2018}
Teeter C, Iyer R, Menon V, Gouwens N, Feng D, Berg J, Szafer A, Cain N, Zeng H,
  Hawrylycz M, Koch C, Mihalas S (2018).
\newblock \enquote{Generalized Leaky Integrate-and-Fire Models Classify
  Multiple Neuron Types.}
\newblock \emph{Nature Communications}, \textbf{9}(1), 1--15.
\newblock \doi{10.1038/s41467-017-02717-4}.

\bibitem[{Thaben and Westermark(2014)}]{Thaben+Westermark:2014}
Thaben PF, Westermark PO (2014).
\newblock \enquote{Detecting Rhythms in Time Series with RAIN.}
\newblock \emph{Journal of Biological Rhythms}, \textbf{29}(6), 391--400.
\newblock \doi{10.1177/0748730414553029}.

\bibitem[{Wickham(2016)}]{Wickham:2016}
Wickham H (2016).
\newblock \emph{\textbf{ggplot2}: Elegant Graphics for Data Analysis}.
\newblock Springer-Verlag New York.
\newblock ISBN 978-3-319-24277-4.
\newblock \urlprefix\url{https://ggplot2.tidyverse.org}.

\bibitem[{Zhang \emph{et~al.}(2014)Zhang, Lahens, Ballance, Hughes, and
  Hogenesch}]{Zhang+Lahens+Ballance+Hughes+Hogenesch:2014}
Zhang R, Lahens NF, Ballance HI, Hughes ME, Hogenesch JB (2014).
\newblock \enquote{A Circadian Gene Expression Atlas in Mammals: Implications
  for Biology and Medicine.}
\newblock \emph{Proceedings of the National Academy of Sciences},
  \textbf{111}(45), 16219--16224.
\newblock \doi{10.1073/pnas.1408886111}.

\end{thebibliography}

\end{document}